\documentclass[floatfix, nofootinbib, aps, twocolumn, preprintnumbers]{revtex4-1}
\usepackage{diagbox}                                     % Pour faire des tableaux slash?s
\usepackage{amsmath}                                     % Pour les maths
\usepackage{amsfonts}                                    % Pour les maths
\usepackage{amssymb}                                     % Lettres pour ensembles math?matiques
\usepackage{mathrsfs}                                    % Fonte "manuscrite" pour les maths
\usepackage[latin9]{inputenc}                            % G?re l'encodage
\usepackage[english]{babel}                               % G?re la langue latex
\usepackage[T1]{fontenc}                                 % G?re l'encodage
\usepackage{setspace}                                    % Permet de modifier l'espacement automatique
\usepackage{enumitem}                                    % Pour faire des ?num?rations
\usepackage{geometry}                                    % Pour g?rer les marges
\usepackage{color}

\usepackage{graphicx}                                    % Pour ins?rer des images
\usepackage{subfigure}                                   % Pour faire des sous figures

\newcommand{\tn}{\textnormal}

\def\jnl@style{\it}
\def\aaref@jnl#1{{\jnl@style#1}}
\def\aaref@jnl#1{{\jnl@style#1}}

\def\aj{\aaref@jnl{AJ}}                   % Astronomical Journal
\def\araa{\aaref@jnl{ARA\&A}}             % Annual Review of Astron and Astrophys
\def\apj{\aaref@jnl{ApJ}}                 % Astrophysical Journal
\def\apjl{\aaref@jnl{ApJ}}                % Astrophysical Journal, Letters
\def\apjs{\aaref@jnl{ApJS}}               % Astrophysical Journal, Supplement
\def\ao{\aaref@jnl{Appl.~Opt.}}           % Applied Optics
\def\apss{\aaref@jnl{Ap\&SS}}             % Astrophysics and Space Science
\def\aap{\aaref@jnl{A\&A}}                % Astronomy and Astrophysics
\def\aapr{\aaref@jnl{A\&A~Rev.}}          % Astronomy and Astrophysics Reviews
\def\aaps{\aaref@jnl{A\&AS}}              % Astronomy and Astrophysics, Supplement
\def\azh{\aaref@jnl{AZh}}                 % Astronomicheskii Zhurnal
\def\baas{\aaref@jnl{BAAS}}               % Bulletin of the AAS
\def\jrasc{\aaref@jnl{JRASC}}             % Journal of the RAS of Canada
\def\memras{\aaref@jnl{MmRAS}}            % Memoirs of the RAS
\def\mnras{\aaref@jnl{MNRAS}}             % Monthly Notices of the RAS
\def\pra{\aaref@jnl{Phys.~Rev.~A}}        % Physical Review A: General Physics
\def\prb{\aaref@jnl{Phys.~Rev.~B}}        % Physical Review B: Solid State
\def\prc{\aaref@jnl{Phys.~Rev.~C}}        % Physical Review C
\def\prd{\aaref@jnl{Phys.~Rev.~D}}        % Physical Review D
\def\pre{\aaref@jnl{Phys.~Rev.~E}}        % Physical Review E
\def\prl{\aaref@jnl{Phys.~Rev.~Lett.}}    % Physical Review Letters
\def\pasp{\aaref@jnl{PASP}}               % Publications of the ASP
\def\pasj{\aaref@jnl{PASJ}}               % Publications of the ASJ
\def\qjras{\aaref@jnl{QJRAS}}             % Quarterly Journal of the RAS
\def\skytel{\aaref@jnl{S\&T}}             % Sky and Telescope
\def\solphys{\aaref@jnl{Sol.~Phys.}}      % Solar Physics
\def\sovast{\aaref@jnl{Soviet~Ast.}}      % Soviet Astronomy
\def\ssr{\aaref@jnl{Space~Sci.~Rev.}}     % Space Science Reviews
\def\zap{\aaref@jnl{ZAp}}                 % Zeitschrift fuer Astrophysik
\def\nat{\aaref@jnl{Nature}}              % Nature
\def\iaucirc{\aaref@jnl{IAU~Circ.}}       % IAU Cirulars
\def\aplett{\aaref@jnl{Astrophys.~Lett.}} % Astrophysics Letters
\def\apspr{\aaref@jnl{Astrophys.~Space~Phys.~Res.}}
\def\bain{\aaref@jnl{Bull.~Astron.~Inst.~Netherlands}} 
\def\fcp{\aaref@jnl{Fund.~Cosmic~Phys.}}  % Fundamental Cosmic Physics
\def\gca{\aaref@jnl{Geochim.~Cosmochim.~Acta}}   % Geochimica Cosmochimica Acta
\def\grl{\aaref@jnl{Geophys.~Res.~Lett.}} % Geophysics Research Letters
\def\jcp{\aaref@jnl{J.~Chem.~Phys.}}      % Journal of Chemical Physics
\def\jgr{\aaref@jnl{J.~Geophys.~Res.}}    % Journal of Geophysics Research
\def\jqsrt{\aaref@jnl{J.~Quant.~Spec.~Radiat.~Transf.}}
\def\memsai{\aaref@jnl{Mem.~Soc.~Astron.~Italiana}}
\def\nphysa{\aaref@jnl{Nucl.~Phys.~A}}   % Nuclear Physics A
\def\physrep{\aaref@jnl{Phys.~Rep.}}   % Physics Reports
\def\physscr{\aaref@jnl{Phys.~Scr}}   % Physica Scripta
\def\planss{\aaref@jnl{Planet.~Space~Sci.}}   % Planetary Space Science
\def\procspie{\aaref@jnl{Proc.~SPIE}}   % Proceedings of the SPIE

\begin{document}

\title{Rotating proto-neutron stars:
spin evolution, maximum mass and I-Love-Q relations.}
\author{Gr\'egoire Martinon$^{1,2}$}
\author{Andrea Maselli$^{1,3}$} %\email{andrea.maselli@roma1.infn.it}
\author{Leonardo Gualtieri$^{1,3}$}%\email{Leonardo.Gualtieri@roma1.infn.it}
\author{Valeria Ferrari$^{1,3}$} %\email{Valeria.Ferrari@roma1.infn.it}

\affiliation{
$^1$  Dipartimento di Fisica, Sapienza, Universit\`a  di Roma, P.le A. Moro 2, 
00185 Roma, Italy\\
$^2$ D\'epartement de Physique, \'Ecole Normale Sup\'erieure de Cachan, 
61 Avenue du Pr\'esident Wilson, 94235 Cachan, France\\
$^3$ Sezione INFN Roma1, P.le .A. Moro 2, 00185, Roma, Italy
}

\newcounter{subeq}
\renewcommand{\thesubeq}{\theequation\alph{subeq}}
\newcommand{\newsubeqblock}{\setcounter{subeq}{0}\refstepcounter{equation}}
\newcommand{\mysubeq}{\refstepcounter{subeq}\tag{\thesubeq}}

%--------------------------------------------------------------------------------

\date{\today}

\begin{abstract}

Shortly after its birth in a gravitational collapse, a proto-neutron
star enters in a phase of quasi-stationary evolution characterized by
large gradients of the thermodynamical variables and intense neutrino
emission.
In few tens of seconds the gradients smooth out while the star
contracts and cools down, until it becomes a  neutron star.
In this paper we study this phase of the proto-neutron star life
including rotation, and employing finite temperature equations of
state.

We model the evolution of the rotation rate, and determine the
relevant quantities characterizing the star.
Our results show that an isolated neutron star cannot reach, at the
end of the evolution, the maximum values of mass and
rotation rate allowed by the zero-temperature equation of state.
Moreover, a mature  neutron star evolved in isolation
cannot rotate too rapidly, even if it is born from a
proto-neutron star rotating at the mass-shedding limit.
We also show that the I-Love-Q relations are violated in the first
second of life, but they are satisfied as soon as the
entropy gradients smooth out.
\end{abstract}

\maketitle

\section{Introduction \label{secI}}

Soon after its birth in a gravitational collapse,
a  proto-neutron star (PNS) is a hot and rapidly evolving object.
The initial evolution, which lasts tens to a few hundreds of
milliseconds, is characterized by strong
instabilities of different nature.
Then the PNS life becomes less hectic, and the star slowly
evolves over timescales of the order of days to several thousand years, to reach the
state of a mature, cold neutron star (NS).

The parameters of a newborn PNS carry the imprint of the supernova 
explosion mechanism, which is still quite poorly understood. 
For instance, it is presently not clear how fast can a PNS initially rotate,
an information which is instrumental  to model the subsequent evolution of
young pulsars.

To study this problem, two main approaches have
been followed in the literature. 
The first is based on modelling the stellar evolution of the massive
progenitor (typically $M\sim 10-30\,M_\odot$)
\cite{Heger:2004qp}, and on numerical simulations of the collapse and
of the supernova explosion.  These simulations are performed either in Newtonian theory
\cite{Wongwathanarat:2012zp} or in full general relativity
\cite{2012PhRvD..86b4026O,Ott:2012mr,Kuroda:2012nc}, and extend up to  $\sim100-1000$ ms after the bounce.
These studies  indicate that,  at this stage, the minimum 
rotation period of a PNS should range from few to $\sim10$ milliseconds. 
However, after the bounce several  processes  and
instabilities are likely to spin-down the PNS 
(see e.g. \cite{2006ApJS..164..130O,2012PhRvD..86b4026O}
and references therein). These will be briefly discussed below.

A second approach is based on astrophysical observations of
young pulsars \cite{FaucherGiguere:2005ny,Popov:2012ng}. In these works, the
initial spin rate is inferred from the spin rates of observed pulsars,
assuming that the star spins down according to the standard magneto-dipole rule,
and that no other dissipative mechanisms are effective. In \cite{Popov:2012ng} the
authors, extending the work of  \cite{FaucherGiguere:2005ny},
found that the initial spin period of a set of 30 young pulsars has a
roughly flat distribution, from $\sim 10$ ms to some hundreds of ms, with most
of the pulsars having periods $\gtrsim100$ ms. This approach
is appropriate to trace back the rotation rate up to some ``initial'' time when
the star became a neutron star, and dynamical processes were no longer
effective. 
Thus, the two approaches cover two non overlapping parts of the stellar
evolution. 

However, something very interesting happens in between:
after the  turbulent phase which characterizes the first few
hundreds of millisecond after the bounce, the star undergoes a more
quiet,  ``quasi-stationary" phase, which  can
be described as a sequence of equilibrium configurations
\cite{Pons99,pons00,pons01,Mezzacappaetal_2010}. The main macroscopic phenomena
which characterize this part of the stellar life,
are the  contraction from an initial  30-40 km radius to a final
radius of 10-15 km,  the star cooling, and the smoothing of  
the intense temperature-entropy  gradients in the interior.
To set an approximate timescale, we can say that this phase starts $\sim 200-500$ ms
after bounce and lasts for  about one minute.
Neutrino processes are obviously very important during this time \cite{Pons99}.

As the star contracts and cools down, its rotation rate changes.
Therefore, modelling this part of the PNS life is important to
understand how to match the results of the studies on stellar evolution 
and supernova explosion, with the rotation rates of young
pulsars observed today.

We shall indicate the three main phases of the stellar evolution as:
{\it phase 1}, involving collapse and post-bounce up to a few hundred
ms; {\it phase 2}, when
the evolution is quasi-stationary, and {\it phase 3} when, having reached
approximately its final radius, the star quietly cools down, and its
spin rate decreases  due to magneto-dipole
emission and possibly secular instabilities.

In this paper we are  interested in modelling the
{\it phase 2}.  Before going into the details of the modelling we would like
to briefly discuss whether the numerous phenomena and instabilities,
which can be active during the stellar evolution, may affect the {\it phase 2}.
%%%%%%%%%%%%%%%%%%%%%%%%%%%%%%%%%%%

\begin{itemize}
\item Dynamical instabilities (e.g. bar-mode,
magneto-hydrodynamical) act on a time-scale of the order of
the rotation period (for a detailed discussion, see
\cite{2003CQGra..20R.105A,2006ApJS..164..130O,2013PhRvD..88j4028F}).
They can only set in when the rotation period is very close to the
mass-shedding limit ($\sim 1-10$ ms); therefore,  they are expected to
be suppressed before {\it phase 2} sets in.

\item Secular instabilities (e.g. r-mode, f-mode instabilities) act on
timescales ranging  from $\sim1\,$s to several years; they become
effective at temperatures $T\lesssim10^{10}$ K, typically reached after
the first minute of life of the PNS
\cite{2001IJMPD..10..381A,2013PhRvD..88j4028F,
2003CQGra..20R.105A}. In {\it phase 2}, the star is
too hot ($T\gtrsim10^{11}$ K) for secular instabilities to be effective
(at least, excluding the case of quark stars or stars having exotic
matter in the core).

\item Neutrinos carry away a significant part of the total mass (up to
$\sim20\%$) \cite{Pons99}, and  of the PNS angular momentum (up to
$\sim40\%$ \cite{2004IAUS..218..3J}).  Neutrino processes are most effective in {\it
phase 1}, during which most of the mass and angular momentum 
 loss occurs, but it may not be
negligible in {\it phase 2} as well. This point will be further
discussed in the following.

\item Strong magnetic winds may significantly spin down a rapidly
rotating PNS,  for surface fields as high as  $\sim10^{15}-10^{16}$ G
\cite{Heger:2004qp,2006ApJS..164..130O}.

\item It has recently been suggested \cite{2007Natur.445...58B} that due
to the standing accretion shock instability (SASI), accretion of matter
surrounding the PNS during the first second after the bounce can
significantly spin-up the PNS. However, subsequent, more accurate
numerical simulations found that this process is not effective
\cite{Rantsiou:2010md}.
\end{itemize}
The above listed processes are not fully understood; however, based on 
current  literature, we shall assume that they are not  effective in
{\it phase 2},
i.e. between $t\sim0.2-0.5$ s and $t\sim 1$ minute after bounce, with the only
exception of mass and angular momentum loss through neutrino emission.
We will not include in our study the effects of
extremely strong magnetic fields.

%%%%%%%%%%%%%%%%%%%%%%%%%%%%%%%%%%%%%%%

In the literature, fully non linear and approximate  numerical codes have been
developed to find the structure of mature neutron stars,  at the end of
thermodynamical evolution. However, no much has been
done to correctly simulate a hot, rapidly rotating proto-neutron star.
In a series of papers published in the late Sixthies, J.B. Hartle 
developed a perturbative approach  which allows to expand Einstein's 
equations for the structure of a rotating star at different orders in the
angular velocity. The equations were derived  assuming a
barotropic equation of state (EoS), appropriate to describe a cold, old
neutron star. Since we are interested in hot, newly born proto-neutron
stars, in this paper we generalize this approach to that case,
at third order in the rotation rate,
and solve the equations numerically using some EoS which have
been proposed in the literature to describe hot proto-neutron stars.

We will compute the mass, radius, moment of inertia and quadrupole
moment for stellar sequences having fixed baryonic mass and varying
angular velocity up to mass-shedding.

In addition, we shall determine the evolution of the PNS rotation rate,
while it cools and contracts during {\it phase 2}, under the assumption
that the  angular momentum decreases due to neutrino
emission; we shall employ the heuristic formula proposed in \cite{2004IAUS..218..3J} to
describe the angular momentum carried away by neutrinos.

In recent years some universal relations have been shown   
to exist, which link  the moment of
inertia, the tidal deformability  and the spin-induced
quadrupole moment  of mature neutron stars
\cite{Yagi:2013bca,Yagi:2013awa,Maselli:2013mva,Doneva:2013rha,
Haskell:2013vha,Chakrabarti:2013tca}. They are named ``I-Love-Q''
relations, and have been tested for cold equations of state ($T\lesssim
10^9$ K).
In the last  section of this paper,
we will assess the range of validity of the ``I-Love-Q'' relations
when applied to  newly-born, hot proto-neutron stars.

%%%%%%%%%%%%%%%%%%%%%%%%%%%%%%%%%%%%%%%%%%
\section{Perturbative approch to rotating stars extended to 
non-barotropic EoS \label{secII}}
%%%%%%%%%%%%%%%%%%%%%%%%%%%%%%%%%%%%%%%%%%
According to the perturbative approach introduced  by Hartle
\cite{Hartle67b}
the metric describing a rotating star can be found by perturbing 
the spherical non-rotating metric in powers
of the angular velocity $\Omega$, expandind in 
Legendre polynomials. The resulting metric up to third order terms in
$\Omega$ is

\begin{eqnarray}
\nonumber ds^2 =& -&e^{\nu(r)} [ 1 + 2h_0(r) + 2h_2(r)P_2(\mu) ] dt^2 \\
\nonumber      &+& e^{\lambda (r)}\left[ 1 + \frac{ 2m_0(r) +
2m_2(r)P_2(\mu) }{ r - 2M(r) } \right]dr^2 \\
               &+& r^2 [ 1 + 2k_2(r)P_2(\mu) ] \times [d\theta^2 +\sin^2\theta \{ d\phi\\
               \label{ds2}
\nonumber   &-&  [\omega (r)
+ w_1(r) + w_3(r)P^{'}_3(\mu)]dt \}^2 ]
\end{eqnarray}
\noindent where  $\mu=\cos\theta$ and $P_n(\mu)$ is the Legendre
polynomial of order $n$, the prime denoting the derivative with respect
to $\mu$. The functions $\nu$, $M$ and $\lambda$ are those commonly used 
to describe the non-rotating stars in
the TOV equations. The function $\omega$ is of order
$\Omega$ and is responsible for the dragging of inertial frames; it also
determines the lowest order contribution to 
the star angular momentum.  The
functions $h_0$, $m_0$ and $h_2$, $m_2$, $k_2$ are all of order
$\Omega^2$, the former giving rise to spherical expansion, the latter to
quadrupolar deformation.  Finally, the functions $w_1$, $w_3$ are of
order $\Omega^3$ and are involved in corrections to the angular momentum
as well as to the frame-dragging and mass-shedding limit (see
\cite{Benhar05} for details).

The energy-momentum tensor of the matter composing  the star is :

\begin{equation}
   T^{\mu\nu} = (\mathscr{E} + \mathscr{P})u^{\mu}u^{\nu} + \mathscr{P}g^{\mu\nu}
\end{equation}

\noindent where calligraphic, capital letters denote thermodynamical quantities
(energy density and pressure) in the rotating configuration, $u^{\mu}$
are the components of the 4-velocity of the fluid and $g_{\mu\nu}$ is the metric.

Due to rotation, an element of fluid located
at a given $(r,\theta)$ in the non rotating star, is displaced to
\begin{equation}
\label{displ1}
\bar{r}= r + \xi(r,\theta)~,
\end{equation}
where
\begin{equation}
\label{disp1}
\xi(r,\theta)=\xi_0(r) + \xi_2(r)P_2(\mu) +O(\Omega^4)
\end{equation}
is the Lagrangian displacement, 
and both $\xi_0$ and $\xi_2$ are of order $\Omega^2$.
As a consequence of this displacement, all thermodynamical variables 
experience a local change which is found, following Hartle
(\cite{Hartle67b}, Eq. (71)),  setting to zero the Lagrangian variation of the
considered quantity. For instance, for the pressure
\begin{equation}
\Delta P = \delta P + \frac{dP}{dr} \delta r =0\quad\rightarrow\quad
 \delta P(r,\theta) = - \frac{dP}{dr}\xi(r,\theta).
\end{equation}
Consequently, we can write
\begin{equation}
\label{defdp}
   \delta P(r,\theta) = [\epsilon(r) + P(r)][\delta p_0(r) + \delta
p_2(r)P_2(\mu)],
\end{equation}
where $\epsilon$ and $P$ are energy density and pressure computed
by the TOV equations, and
\begin{equation}
\label{dpxi}
   \delta p_{0,2} = -\xi_{0,2}\left[ \frac{1}{\epsilon + P}\frac{dP}{dr}
\right]~,
\end{equation}
and similarly, for the Eulerian change of the energy density. 
It should be noted that, if the equation of state is barotropic
$\epsilon=\epsilon(P)$, this change can be written as
\begin{equation}
\delta \epsilon= - \frac{d\epsilon}{dr}\xi(r,\theta) =
\frac{d\epsilon}{dP} [\epsilon(r) + P(r)][\delta p_0(r) + \delta
p_2(r)P_2(\mu)]~,\label{defdeps}
\end{equation}
where   $\frac{d\epsilon}{dP} =(d\epsilon/dr)/(dP/dr)$.
However, since we are interested in generalizing  Hartle's
equations to the non-barotropic case,  when
$\epsilon=\epsilon(p,s,Y_i)$, where $s$ is the entropy
per baryon, and $Y_i$ is the number fraction of the $i$-th specie,
we do not introduce the total derivative
$\frac{d\epsilon}{dP}$ in the expression of the mass-energy
density perturbation $\delta\epsilon$, as in Eq.~(\ref{defdeps}). 
We simply write
\begin{equation}
\delta\epsilon=\frac{d\epsilon}{dr}/\frac{dP}{dr}
(\epsilon(r)+P(r))[\delta p_0(r)+\delta p_2(r)P_2(\mu)]\,.
\end{equation}
We remark that in the case of a non-barotropic EoS,
in order to compute the mass-energy density
$\epsilon$ and the total radial
derivative $\frac{d\epsilon}{dr}$ for an assigned value of  pressure,
the entropy and the number fraction profiles, $s(P)$, $Y_i(P)$, must
be specified.

Solving Einstein's equations is equivalent to solve a set of
differential equations for all the perturbation functions defined above.
These equations are summarized in the appendix of \cite{Benhar05}. In
the non-barotropic
case, there are only two differences with respect of the original
derivation of
\cite{Hartle67b,Hartle73}. The first is that, as discussed above,
$d\epsilon/dP$ should
be meant as a short-hand for
$\frac{d\epsilon}{dr}/\frac{dP}{dr}$. Secondly, in 
\cite{Hartle67b} equations
\begin{eqnarray}
\frac{d}{dr}\left(\delta p_0 + h_0 - \frac{\chi^2r^3}{3(r - 2M)}\right)
&=&0\label{const0}\\
\delta p_2 + h_2 - \frac{\chi^2r^3}{3(r - 2M)}  &=& 0\label{const2}
\end{eqnarray}
are obtained by exploiting a first integral of Einstein's equations,
\begin{equation}
{\rm const.} = \frac{\mathscr{E} + \mathscr{P}}{u^t}{\rm exp}
\left( -\int\frac{d \mathscr{E}}{\mathscr{E} + \mathscr{P}} \right)\,.
\end{equation}
To our knowledge, no such first integral exists 
in the non-barotropic case.
However, Eqs.~(\ref{const0}) and (\ref{const2}) are still valid since they
can be shown to follow
from the divergence equation $T^{\mu\nu}_{~~;\nu}=0$.

%&&&&&&&&&&&&&&&&&&&&&&&&&&&&&&&&&&&&&&&&&&&&&&&&&&&&&&&&&&&&&&&&&
\section{Tests of Hartle's procedure versus full GR \label{secIII}}
\begin{figure}[htbp]
\centering
\hspace{-1.3cm}
\subfigure{\includegraphics[width=7.5cm]{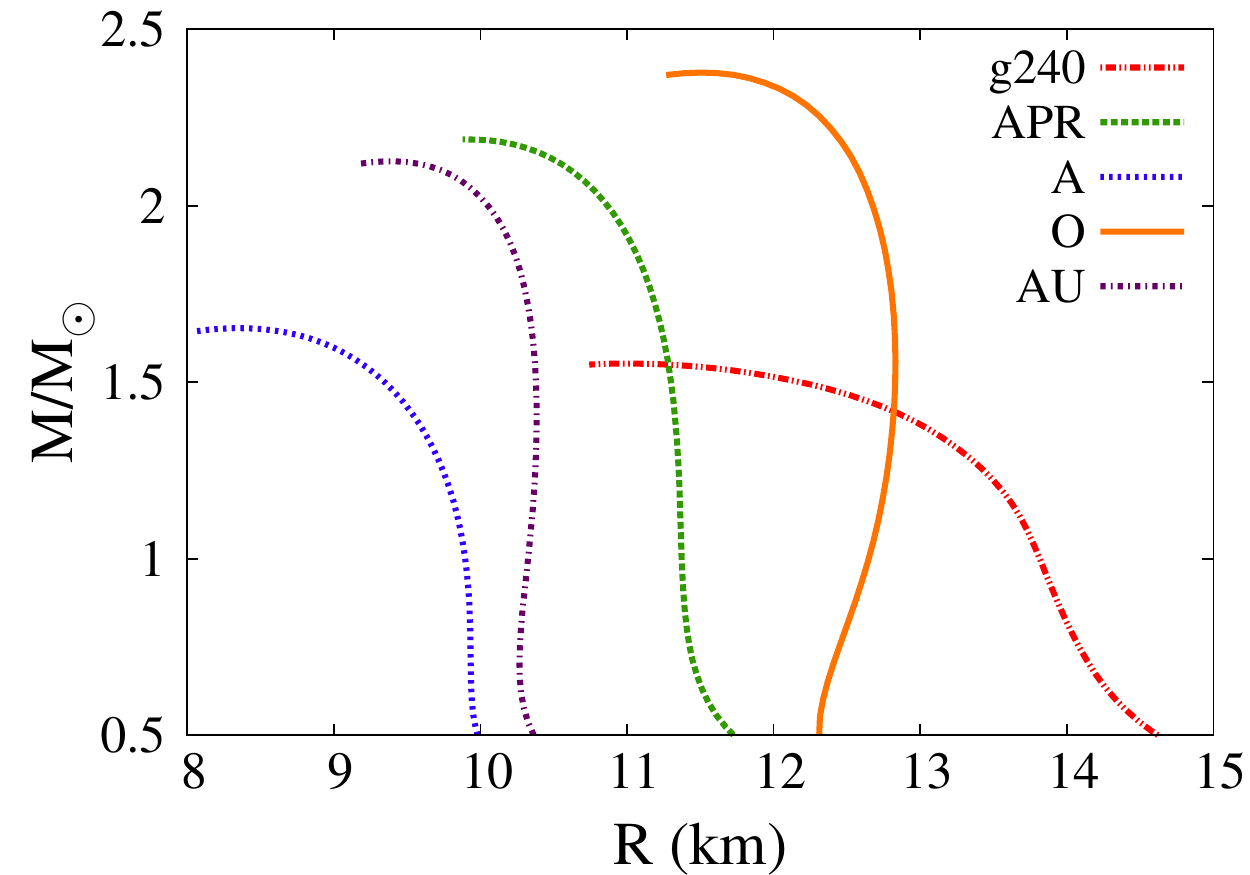}}
\caption{(Color online) Mass-Radius diagram for the cold EoS used to test Hartle's
procedure versus RNS code results.}
\label{FIG1}
\end{figure}

%&&&&&&&&&&&&&&&&&&&&&&&&&&&&&&&&&&&&&&&&&&&&&&&&&&&&&&&&&&&&&&&&&
In order to establish to what extent Hartle's procedure gives an
accurate description of a rotating star, we have  used  five barotropic
EoS appropriate to describe mature neutron stars, 
named \texttt{A}, \texttt{AU}, \texttt{APR}, \texttt{O}, \texttt{g240}, 
(see the Appendix for details), whose mass-radius relations are shown in Fig.~\ref{FIG1}.
The reason why we choose these EoS is that they span a large range of
stellar compactness.
Tests of Hartle's procedure for cold NS,
have been done in several papers in the past, to
establish the relevance of terms of order $n+1$ in the
rotation rate with respect to those of order $n$ \cite{Benhar05}, or to
compare the results of the numerical integration of Hartle's equations
with those of fully relativistic, non linear codes
\cite{Cook:1993qr,Berti04,Berti05}.  
Since we are interested in  constructing sequences of 
newly born, rotating proto-neutron stars, which 
are less dense than the cold NS which form at
the end of the evolution, we want to understand in particular how the
accuracy of Hartle's procedure depends on the stellar compactness.
Therefore we have  compared the results we obtain integrating Hartle's equations
for  stars with fixed baryonic mass and different EoS,
with those we find using the open source code RNS \cite{rns}, which integrates the fully
nonlinear equations of stellar structure.
We would like to stress that this test is possible only for cold EoS, since
fully non linear codes for rapidly rotating, hot proto-neutron stars are not
publicly available. 
The quantities we compare are the mass, radius, moment of inertia. We do
not compare the quadrupole moment, because  the computation of this
quantity in the public version of RNS is affected by a systematic error
\cite{pappas_apostolatos2012}.
%%%%%%%%%%%%%%%%%%%%%%%%%%%%%%%%%%%%%%%%%%%%%%%%%%%%%%%%%%%%%%%%%%
%%%%%%%%%%%%%%%%%%%%%%%%%%%%%%%%%%%%%%%%%%%%%%%%%%%%%%%%%%%%%%%%%%
Details on this comparison are given in the Appendix, here we summarize
the results. 

It is known that the mass-shedding rotation rate $\nu_{\tn{ms}}$ is
systematically
overestimated by the perturbative approach \cite{Benhar05}. Therefore, we extend
the comparison up to $\nu_{\tn{ms}}$ evaluated by the RNS code, and
normalize the rotation rate to that value.
We choose two values of the baryonic mass,
$M_\tn{b}=1.55~M_\odot$ and $M_\tn{b}=2.2~M_\odot$. For $M_\tn{b}=1.55~M_\odot$
we consider the EoS \texttt{A}, \texttt{APR} and \texttt{g240}, whereas for
$M_\tn{b}=2.2~M_\odot$ we consider the EoS \texttt{AU}, \texttt{APR}, 
and \texttt{O} since the maximum mass of \texttt{A} and \texttt{g240} 
is smaller than this value.

The results of the Hartle-versus-RNS comparison are given in
Fig.~\ref{FIG9} and in Table~\ref{table7} of the Appendix.
These show that the two approaches produce values of the mass,
equatorial radius (circumferential), and moment of inertia in good agreement (relative
difference $\lesssim 5\%$ for rotation rates up to 0.8 of the
mass-shedding limit), quite independently of the stellar compactness, and EoS.
This gives a strong indication for the reliability of Hartle's procedure 
when applied to hot and less dense stars, which is the case  we
are interested in. 

In the next sections we shall apply Hartle's procedure to study the
structure of newly born, rotating proto-neutron stars. 
We shall extend our calculations up to the mass-shedding limit,
keeping in mind the limitations on the accuracy discussed above.
%TTTTTTTTTTTTTTTTTTTTTTTTTTTTTTTTTTTTTTTTTTTTTTTTTTTTTTTTTTTTTTTTTTTT
\begin{table*}
\centering
\begin{tabular}{ccccc|ccccc}
\hline
\hline
\multicolumn{10}{c}{$M_\tn{b} = 1.6 M_{\odot}$}\\
\hline
\texttt{GM3NQ}&$M/M_\odot$ & $R$ (km) & $I/I^*$ & $\nu_{\tn{ms}}$ (Hz)&
\texttt{BS}&$M/M_\odot$ & $R$ (km) & $I/I^*$ & $\nu_{\tn{ms}}$ (Hz)\\
\hline
$t=0.2$ s & 1.58 & 34.35 & 5.33 & 97    & -                 & -    & -     & -    & - \\
$t=0.5$ s & 1.56 & 23.78 & 3.75 & 306   & \texttt{S1S5-032}  & 1.50 & 24.35 & 1.84 & 256 \\
$t=2$ s   & 1.53 & 15.78 & 2.44 & 718   & \texttt{S2S4-032}  & 1.51 & 19.51 & 1.83 & 420  \\
$t=5$ s   & 1.50 & 13.61 & 2.00 & 931   & \texttt{S1S2-023}  & 1.49 & 14.02 & 1.45 & 790   \\
$t=20$ s  & 1.47 & 12.91 & 1.76 & 1010  & \texttt{T0}        & 1.43 & 11.80 & 1.41 & 1060  \\
\hline
\hline
\end{tabular}
\caption{We compare some relevant parameters  of the non rotating
configurations belonging to two models of evolving proto-neutron
stars with the same baryonic mass:  \texttt{GM3NQ} and \texttt{BS}. 
 \texttt{GM3NQ} describes a true evolutionary sequence of a proto-neutron
star, whereas the sequence \texttt{BS} mimics the evolution of a star
with EoS different from \texttt{GM3NQ}, and similar entropy gradients. 
For both sequences we tabulate the gravitational mass, the
radius,  the moment of inertia (normalized to $I^* = 10^{45}$
g$\times$cm$^2$), and  mass-shedding frequency
evaluated using the fit (\ref{MSfitDoneva}).}
\label{table1}
\end{table*}
%TTTTTTTTTTTTTTTTTTTTTTTTTTTTTTTTTTTTTTTTTTTTTTTTTTTTTTTTTTTTTTTTTTTT

%%%%%%%%%%%%%%%%%%%%%%%%%%%%%%%%%%%%%%%%%%%%%%%%%%%%%%%%%%%%%%%%%%
\begin{figure*}[ht]
\centering
\hspace{-1.3cm}
\subfigure{\includegraphics[width= 7.5cm]{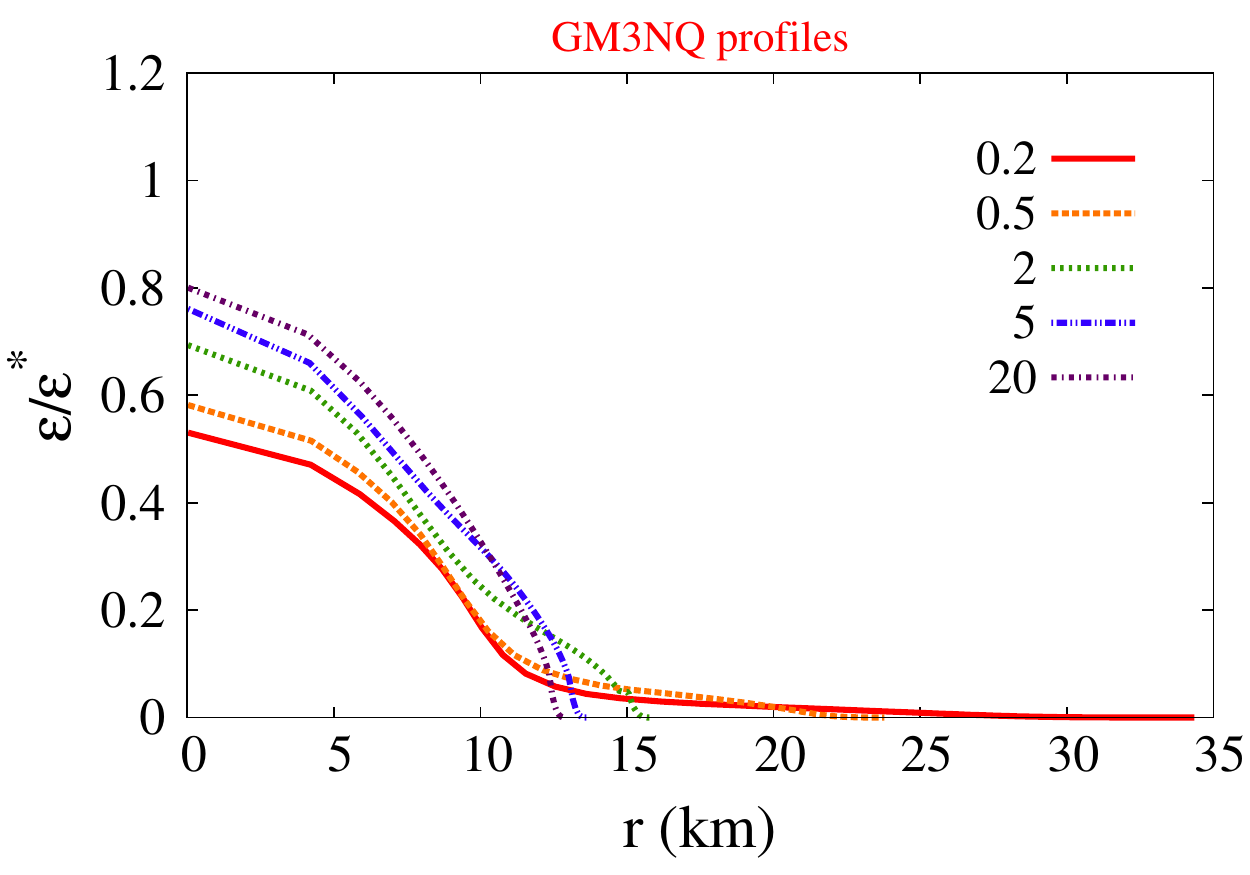}}
\subfigure{\includegraphics[width= 7.5cm]{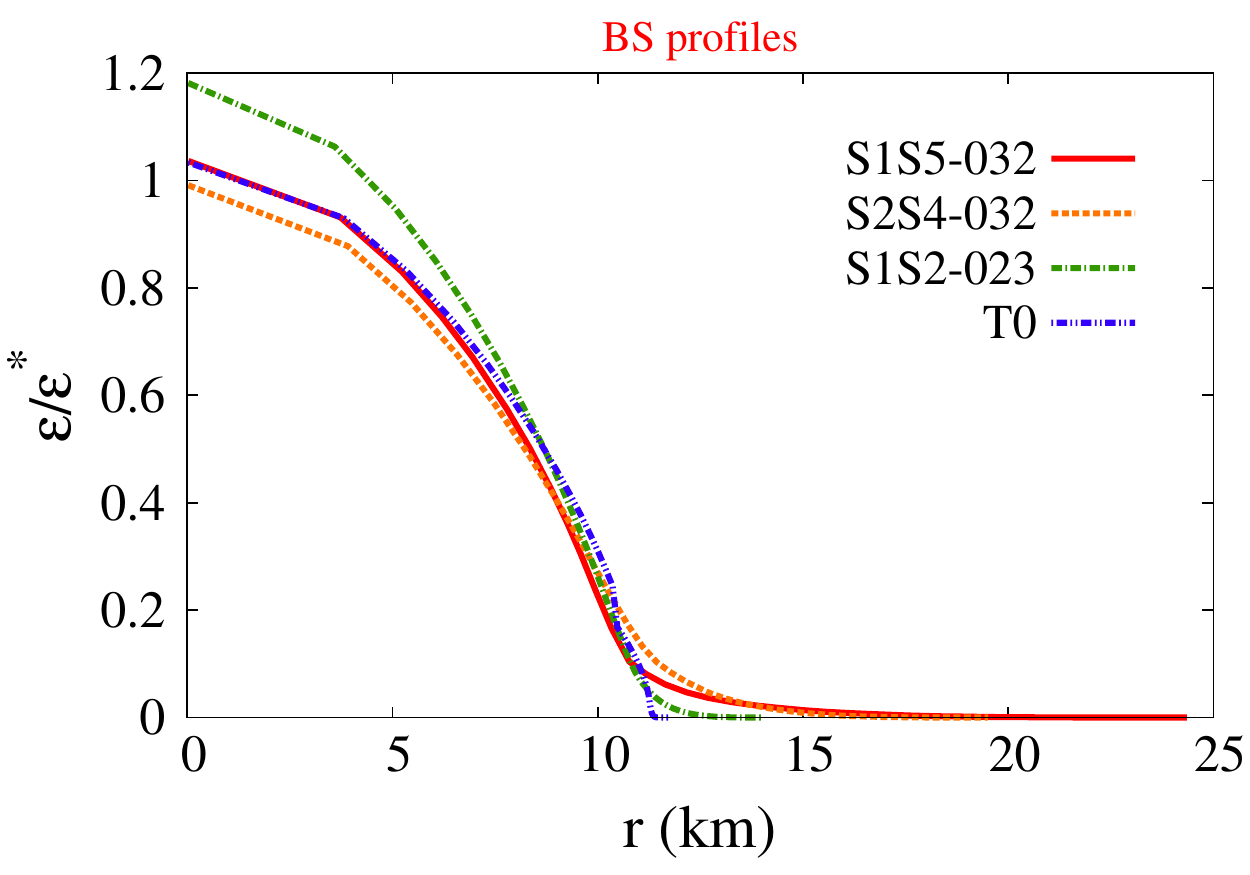}}
\\\noindent
\subfigure{\includegraphics[width= 7.5cm]{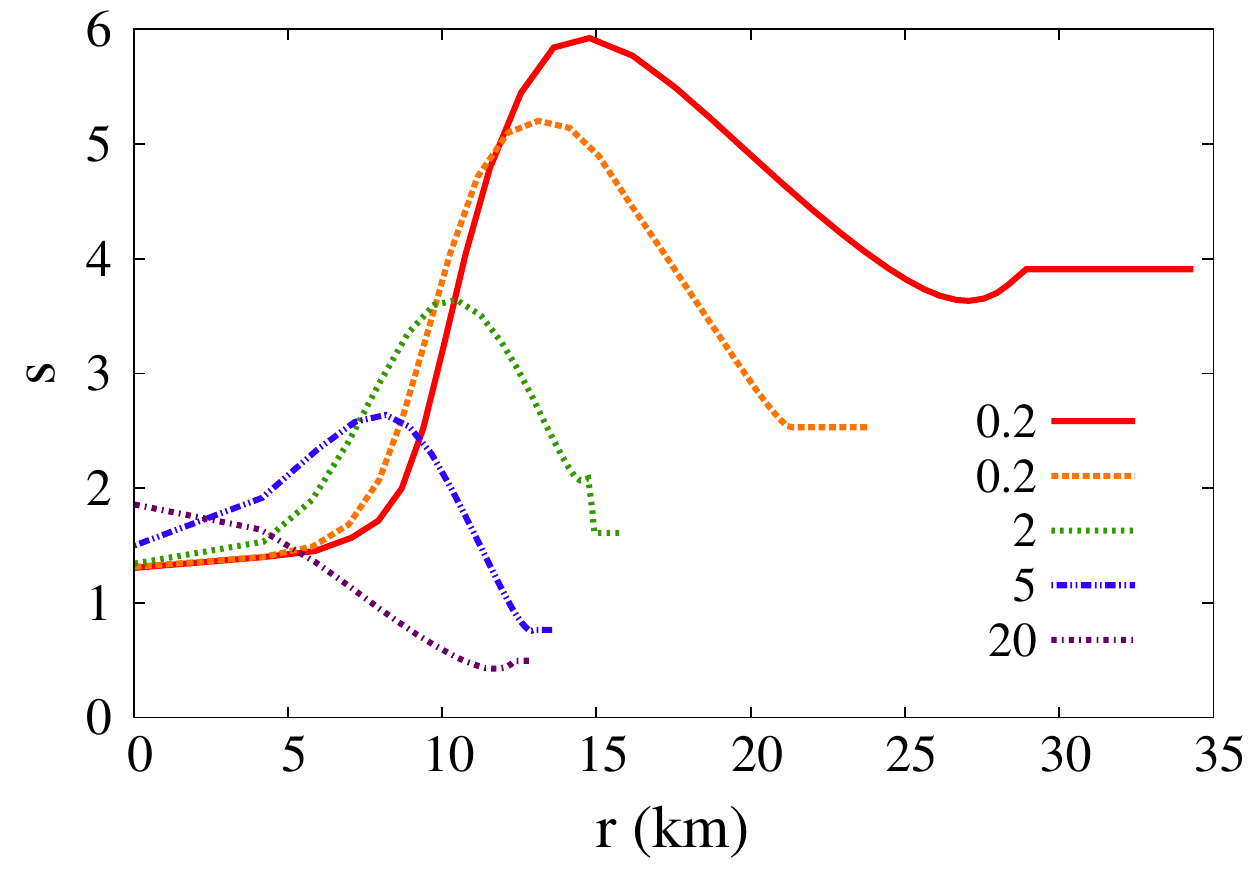}}
\subfigure{\includegraphics[width= 7.5cm]{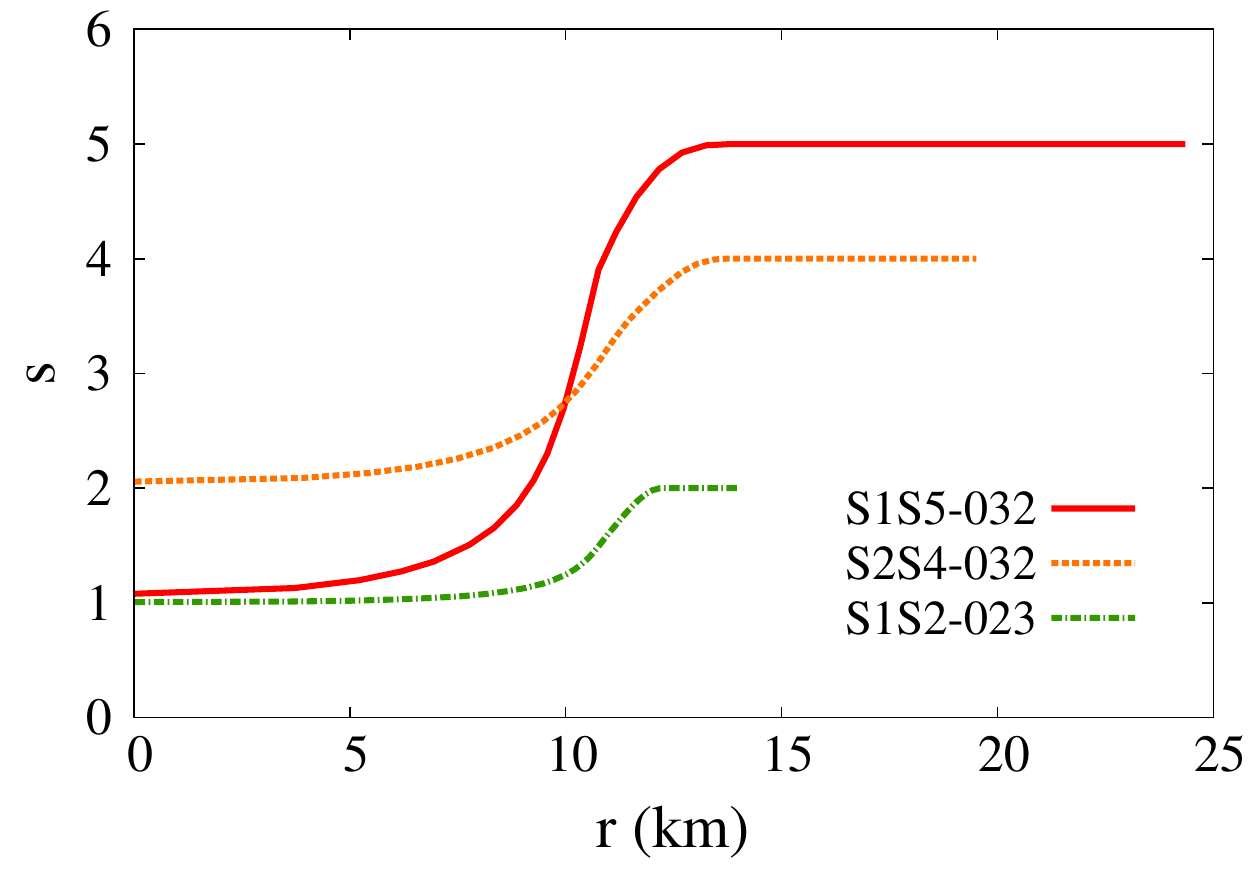}}
\caption{(Color online) The  energy density and entropy per baryon versus radial
distance are plotted for the two families of hot stars \texttt{GM3NQ}
(left)
and \texttt{BS} (right). The energy density (upper panel) is
normalized to the value $\epsilon_*=10^{15}$ g/cm$^3$.
All plots refer to a non rotating star with baryonic mass
$M_\tn{b}=1.6~M_\odot$.
}
\label{FIG2}
\end{figure*}
%%%%%%%%%%%%%%%%%%%%%%%%%%%%%%%%%%%%%%%%%%%%%%%%%%%%%%%%%%%%%%%%%%

%%%%%%%%%%%%%%%%%%%%%%%%%%%%%%%%%%%%%%%%%%
\section{Models of evolving PNS \label{secIV}}
%%%%%%%%%%%%%%%%%%%%%%%%%%%%%%%%%%%%%%%%%%
Numerical simulations of the early evolution of newly born proto-neutron
stars have shown that a few tenths of seconds after the bounce
which follows a supernova explosion, the  evolution of the star can be
considered as ``quasi-stationary", i.e. it can
be described by  a sequence of equilibrium configurations
\cite{Pons99,pons00,pons01,Mezzacappaetal_2010}. 
The main feature of this quasi-stationary phase
is that,  starting from an initial configuration
characterized by a low-entropy core
and a high-entropy envelope, due to neutrino processes the
entropy gradient gradually smoothes out, while the star
progressively cools down, and eventually the overall entropy decreases.

To study how the structure of an evolving  PNS depends on the rotation
rate, we use two sets of non-barotropic, hot stellar models,
based on two different equations of state of baryonic matter.
The first is a sequence of mass-energy, pressure, lepton fraction and
entropy profiles (``profiles'' in short),
indicated as \texttt{GM3NQ} to hereafter,
based on an equation of state obtained within a finite-temperature,
field-theoretical model solved at the mean field level
\cite{Pons99,pons00,pons01}.
These profiles were found by solving the  relativistic equations of
neutrino transport and nucleon-meson coupling, assuming a spherical
spacetime background, and a NS baryonic mass $M_{\tn{b}}=1.6
M_{\odot}$, during the first minute of the
stellar life. Thus, this is a true evolutionary sequence, and it refers to a
unique baryonic mass.
These models have been used to compute how
the oscillation frequencies change as the star cools and contracts in
\cite{Ferrari03A} (Model A), \cite{Ferrari03B}, and in
\cite{Gualtieri04} to
study how thermal diffusion affects the oscillation frequencies.

The second set of non-barotropic, hot profiles we shall use, tagged  as
\texttt{BS}, are based on a  microscopic EoS obtained within the
Brueckner-Hartree-Fock  nuclear many-body approach extended to the
finite temperature regime within the Bloch-De Dominicis formalism (see
\cite{book,bloch,Burgio10,Burgio11} and references therein for explicit
details).  These models can be used to
``mimic'' a PNS evolution, since each profile is characterized by
an entropy gradient which reproduces the main features of the
quasi-stationary evolution described above. Thus, even though they are not
obtained as a result of a dynamical simulation, they  can be
 considered as snapshots of different stages of the evolution.
The \texttt{BS} profiles are specified by three
parameters: the entropy per baryon in the core $s_c$, the entropy per
baryon in the envelope $s_e$, and the leptonic fraction $Y_e$, which is
assumed to be constant throughout the star.
In the following, we will identify the \texttt{BS}
profiles with a particular choice of these three parameters with the
label \texttt{Ss$_{c}$Ss$_{e}$-Y$_{e}$}, as in \cite{Burgio11}. We will use
an additional profile at zero temperature, which will be tagged as
\texttt{T0}.

It should be stressed that the \texttt{BS} profiles can be used to generate
rotating stellar models using Hartle's procedure because, for each value
of the rotation rate we can change the value of the central energy
density and find stellar configurations with assigned baryonic mass.
This is not possible with the profiles \texttt{GM3NQ}, since they
correspond to a fixed value of the baryonic mass and of the central
energy density.

The \texttt{BS} profiles will be used in the next two section to study how the 
structure of a proto-neutron star changes as a function of the rotation
rate.
The evolutionary sequence \texttt{GM3NQ} will be used in  section
\ref{secVII} to test the ``I-Love-Q'' relations.

%%%%%%%%%%%%%%%%%%%%%%%%%%%%%%%%%%%%%%%%%%
\section{Hot and young neutron stars.  \label{secV}}
%%%%%%%%%%%%%%%%%%%%%%%%%%%%%%%%%%%%%%%%%%

In order to study the behaviour of the relevant parameters 
of evolving proto-neutron stars as a function of the rotation rate,
it is useful to know their values for the non rotating configurations
having the same baryonic mass. 
In Table~\ref{table1} we tabulate the gravitational mass, the
radius and  the moment of inertia for a  star 
with baryonic mass $M_\tn{b}=1.6~M_\odot$ belonging to the
profiles \texttt{GM3NQ} and \texttt{BS}.
As explained in section~\ref{secIV},
\texttt{GM3NQ} describes a true evolutionary sequence of a proto-neutron
star, and in the first column we give the time after collapse
at which each quasi-stationary configuration has been computed (see 
\cite{Pons99} and \cite{Ferrari03A} for details). 

The sequence \texttt{BS} mimics the evolution of a star
with entropy gradients similar to those of the \texttt{GM3NQ} sequence.
These profiles are labelled as \texttt{S1S5-032}, \texttt{S2S4-032},
\texttt{S1S2-023}, \texttt{T0}, where the first two numbers indicate the entropy per
baryon respectively in the core and in the envelope, and the last
is the lepton fraction. 
Thus, for instance, \texttt{S1S5-032} corresponds
to the largest entropy gradient between core and envelope,
and a lepton fraction $Y_l=0.32$. 
\texttt{T0} is the zero
temperature, endpoint of the \texttt{BS} sequence.
A comparison of the entropy profile,
mass and radius of this configurations with those of the
\texttt{GM3NQ} sequence indicates that the  \texttt{S1S5-032}
proto-neutron star is similar to that of the  \texttt{GM3NQ}
sequence corresponding to $t=0.5$ s. 

In Fig.~\ref{FIG2} we show the energy density and the entropy profiles
inside the stars belonging to the two families.

As in Table~\ref{table1},
in Table~\ref{table2} we show the stellar parameters
of stars belonging to the \texttt{BS} sequence
with $M_{\tn{b}}=1.8 M_{\odot}$ and $M_{\tn{b}}=2.0 M_{\odot}$.

%TTTTTTTTTTTTTTTTTTTTTTTTTTTTTTTTTTTTTTTTTTTTTTTTTTTTTTTTTTTTTTTTTTTT
\begin{table}[ht]
\centering
\begin{tabular}{ccccc}
\hline
\hline
EoS & $M/M_\odot$ & $R$ (km) & $I/I^*$ & $\nu_{\tn{ms}}$ (Hz)\\
\hline
&\multicolumn{4}{c}{$M_\tn{b} = 1.8 M_{\odot}$} \\
\hline
\texttt{S1S5-032} & 1.65 & 19.38 & 1.80 & 456   \\
\texttt{S2S4-032} & 1.67 & 16.84 & 1.87 & 596   \\
\texttt{S1S2-023} & 1.65 & 13.11 & 1.60 & 938   \\
\texttt{T0}       & 1.59 & 11.62 & 1.59 & 1150  \\
\hline
& \multicolumn{4}{c}{$M_\tn{b} = 2.0 M_{\odot}$} \\
\hline
\texttt{S1S5-032} & 1.79 & 16.41 & 1.85 & 658   \\
\texttt{S2S4-032} & 1.82 & 14.80 & 1.93 & 784   \\
\texttt{S1S2-023} & 1.80 & 12.20 & 1.72 & 1100  \\
\texttt{T0}       & 1.74 & 11.37 & 1.76 & 1240  \\
\hline
\hline
\end{tabular}
\caption{Parameters  of the non rotating
stellar models  \texttt{BS}, tabulated as in Table~\ref{table1}, for
baryonic mass $M_\tn{b} = 1.8 M_{\odot}$ and $M_\tn{b} = 2.0 M_{\odot}$.
}
\label{table2}
\end{table}
%TTTTTTTTTTTTTTTTTTTTTTTTTTTTTTTTTTTTTTTTTTTTTTTTTTTTTTTTTTTTTTTTTTTT

%%%%%%%%%%%%%%%%%%%%%%%%%%%%%%%%%%%%%%%%%%
\subsection{Mass-shedding limit}
\begin{figure}[ht]
\centering
\subfigure{\includegraphics[width = 8cm]{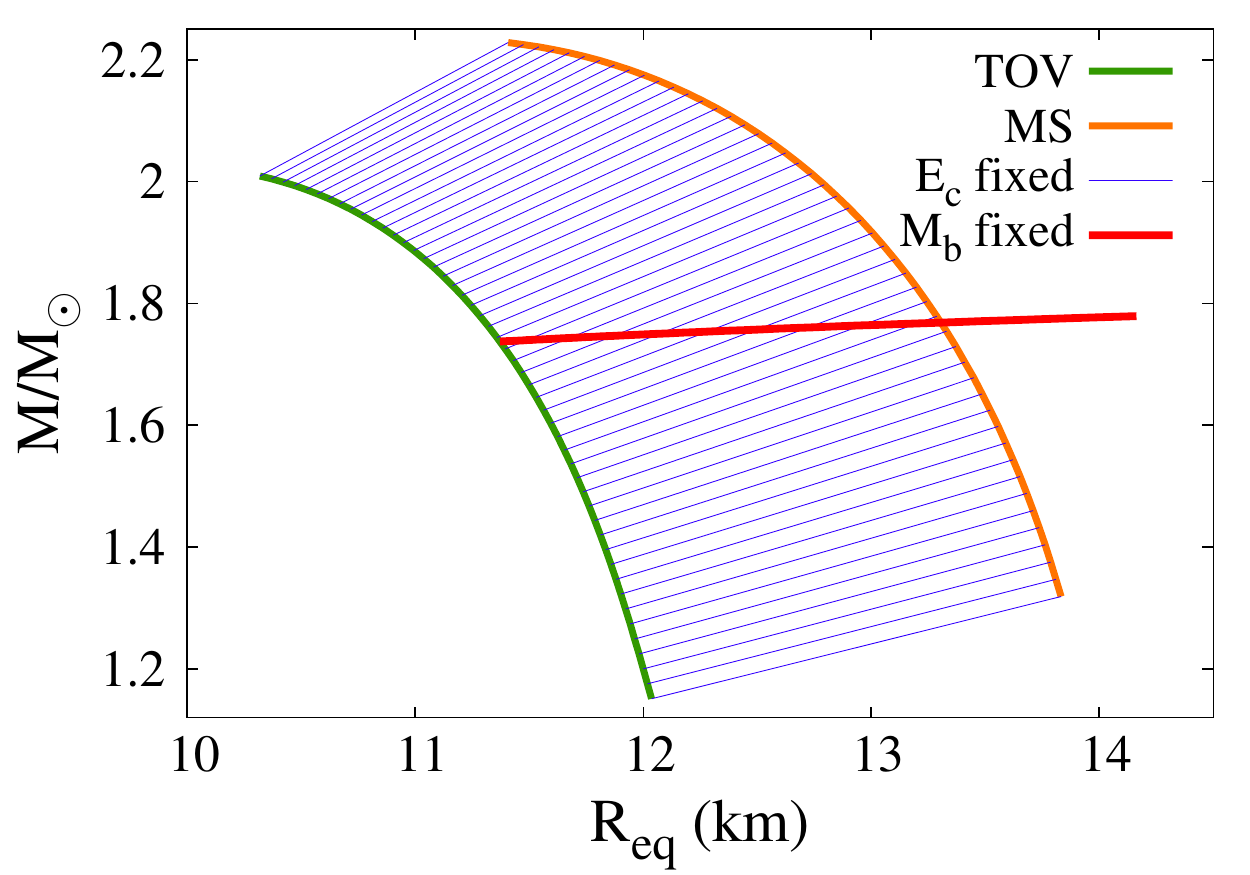}}
\caption{(Color online)
This figure shows, for the profile \texttt{T0},
 how do we compute the mass shedding limit using the
fit (\ref{MSfitDoneva}).
The thin, blue lines are sequences of stellar models with fixed central
energy density $\epsilon_c$ and increasing rotation rate, from zero
(where the blue lines intersect the TOV, zero rotation, green line) up
to mass shedding given by Eq.~(\ref{MSfitDoneva}), represented by the bold,
yellow line.
The thick, nearly horizontal red line is a
sequence at fixed baryonic mass ($M_\tn{b}=2.0 M_\odot$) and increasing rotation rate.
The mass-shedding frequency is given by the intersection of this line
with the bold, yellow line. 
}
\label{FIG3}
\end{figure}

%%%%%%%%%%%%%%%%%%%%%%%%%%%%%%%%%%%%%%%%%%
As we know, Hartle's procedure overestimates the mass-shedding
frequency \cite{Benhar05}. To compute this quantity 
we shall use a different approach, based on fully relativistic
simulations. 
In \cite{Doneva13}, using the RNS code a fit has been proposed 
\begin{equation}
   \nu_{\tn{ms}}(\text{Hz}) = 45862\sqrt{\frac{M_0/M_{\odot}}{(R_0/1
\text{km})^3}} - 189
   \label{MSfitDoneva}
\end{equation}
which relates $\nu_{\tn{ms}}$ to the mass $M_0$ and the radius $R_0$ of
the non-rotating star. According to \cite{Doneva13},
Eq.~(\ref{MSfitDoneva}) estimates $ \nu_{\tn{ms}}$, 
independently of the EoS, with errors not larger than
$\sim 2\%$.  
The fit (\ref{MSfitDoneva}) has been obtained fixing the
central density, and finding the maximum rotation rate as a function of
the gravitational mass and radius given by the TOV equations. 
Since we are interested in studying fixed baryonic mass sequences, to use this
fit we proceed as follows (see Figure~\ref{FIG3}).
We choose a value of  $M_\tn{b}$, and find the corresponding 
TOV-central density $\epsilon_0$.
Then we increase the rotation rate by a small amount reaching, say,
$\Omega_1$,  and find the
mass and radius corresponding to the same $M_\tn{b}$ using Hartle's procedure. 
The central density of this configuration, $\epsilon_1$, is smaller than 
$\epsilon_0$.  The mass and radius  $M_0$ and $R_0$ computed with TOV
for that central density are then inserted into the fit (\ref{MSfitDoneva}) 
to find the corresponding mass-shedding limit. 
If this value is larger than $\Omega_1$, we proceed
further increasing the rotation rate by another small step,
keeping $M_\tn{b}$ fixed and iterate the procedure. If it is
smaller than $\Omega_1$, we stop and say that this is the mass-shedding limit of our
rotating sequence with fixed baryonic mass. The accuracy of our
iterative procedure (using Brent's method)
is of the order of $10^{-5}$, which is smaller than
the estimated accuracy of the fit. 

The values of $\nu_{\tn{ms}}$ evaluated with this procedure for the
sequences \texttt{GM3NQ} and \texttt{BS} are given in the last columns of
Tables~\ref{table1} and \ref{table2}. It is interesting to note that the 
mass-shedding frequences of the initial configuration of each 
sequence are quite low, and  increase as the star cools down and
contracts.
%FFFFFFFFFFFFFFFFFFFFFFFFFFFFFFFFFFFFFFFFFFFFFFFFFFFFFFFFFFFFFFFFFF
%FFFFFFFFFFFFFFFFFFFFFFFFFFFFFFFFFFFFFFFFFFFFFFFFFFFFFFFFFFFFFFFFFF

%FFFFFFFFFFFFFFFFFFFFFFFFFFFFFFFFFFFFFFFFFFFFFFFFFFFFFFFFFFFFFF
\begin{figure*}[ht]
\centering
\hspace{-1.3cm}
\subfigure{\includegraphics[width= 7.5cm]{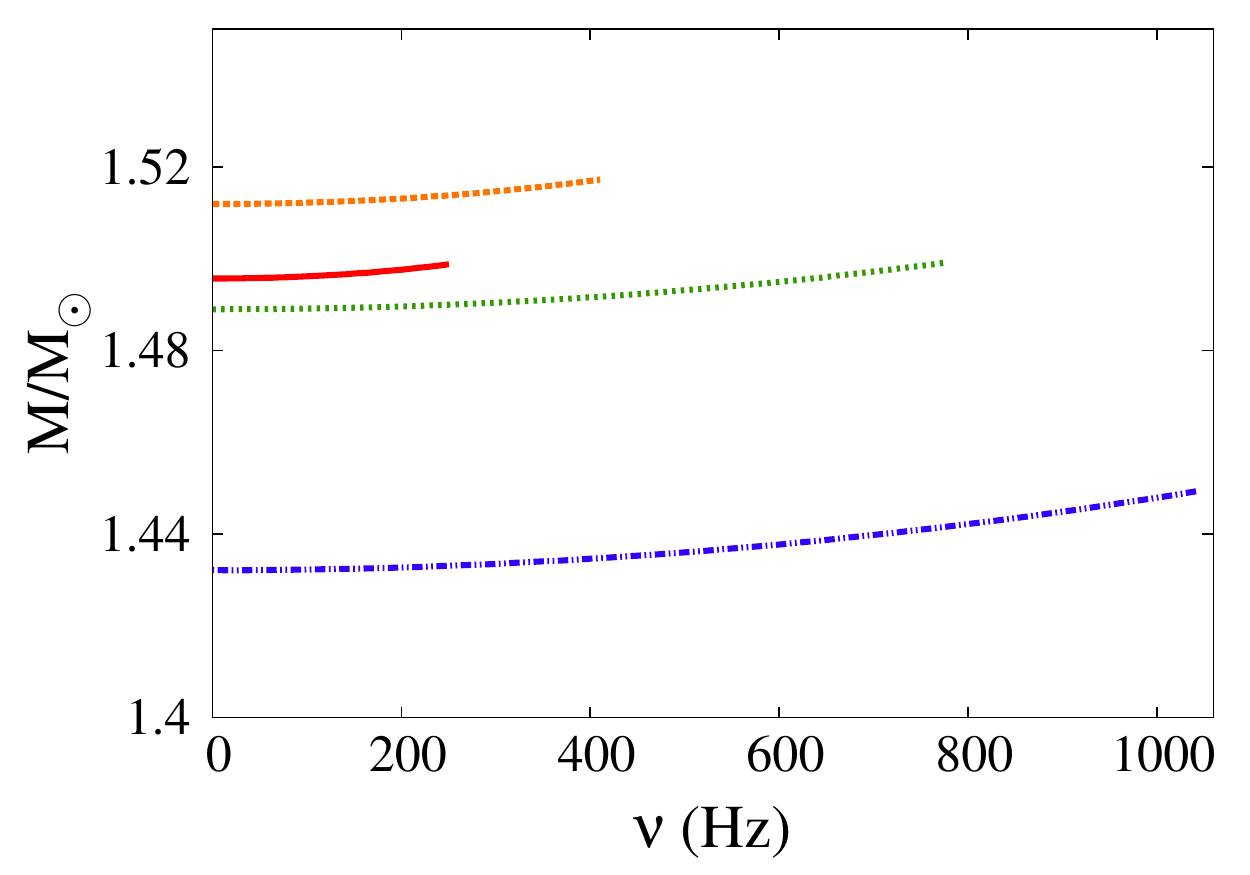}}
\subfigure{\includegraphics[width= 7.5cm]{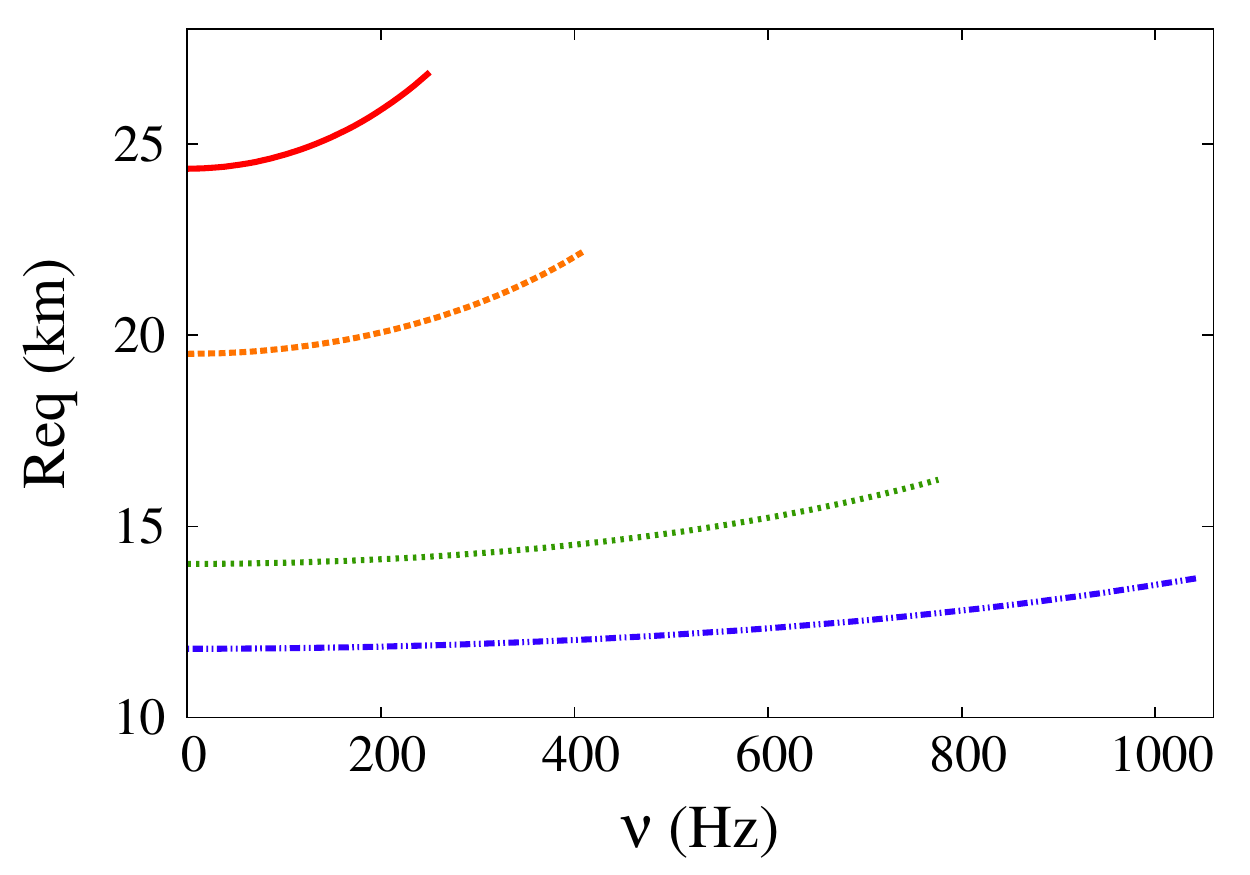}}
\\\noindent
\subfigure{\includegraphics[width= 7.5cm]{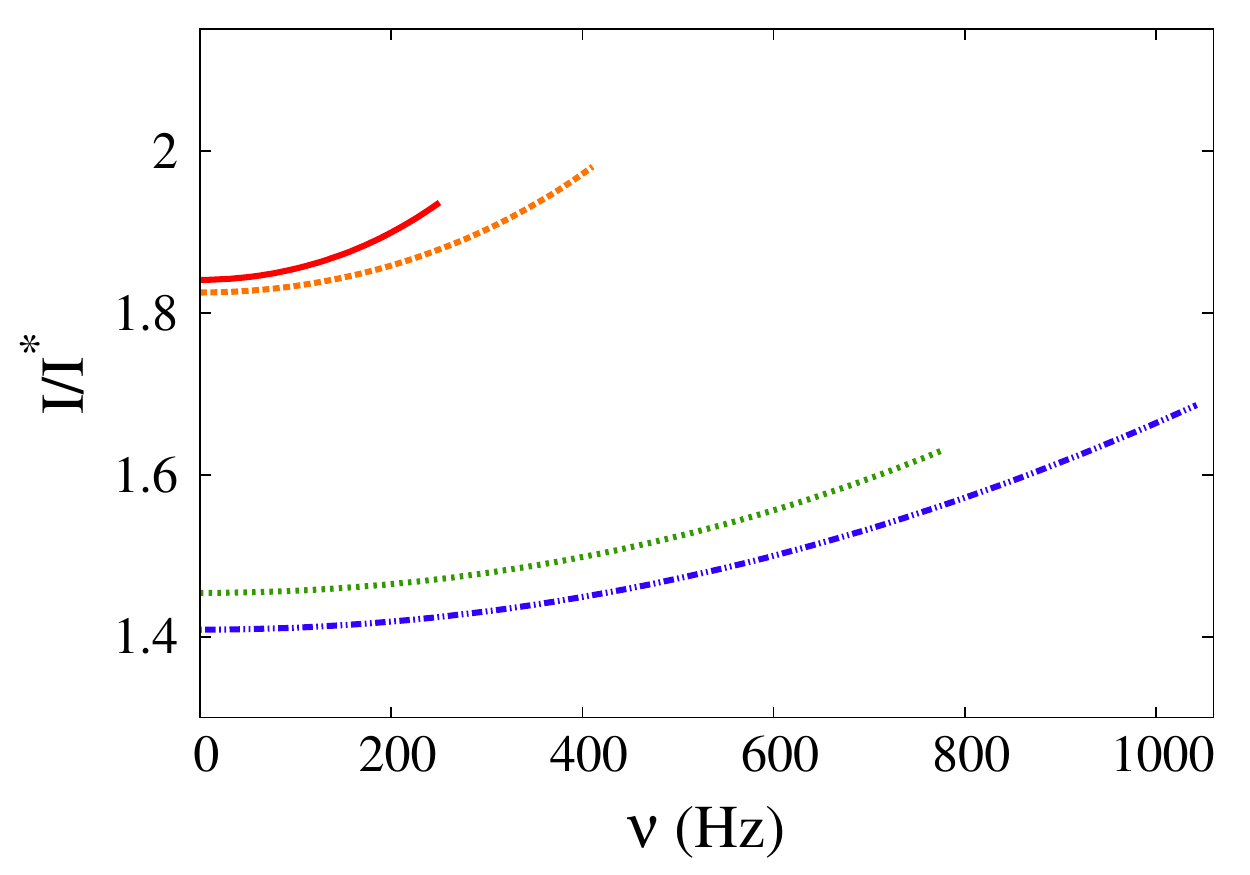}}
\subfigure{\includegraphics[width= 7.5cm]{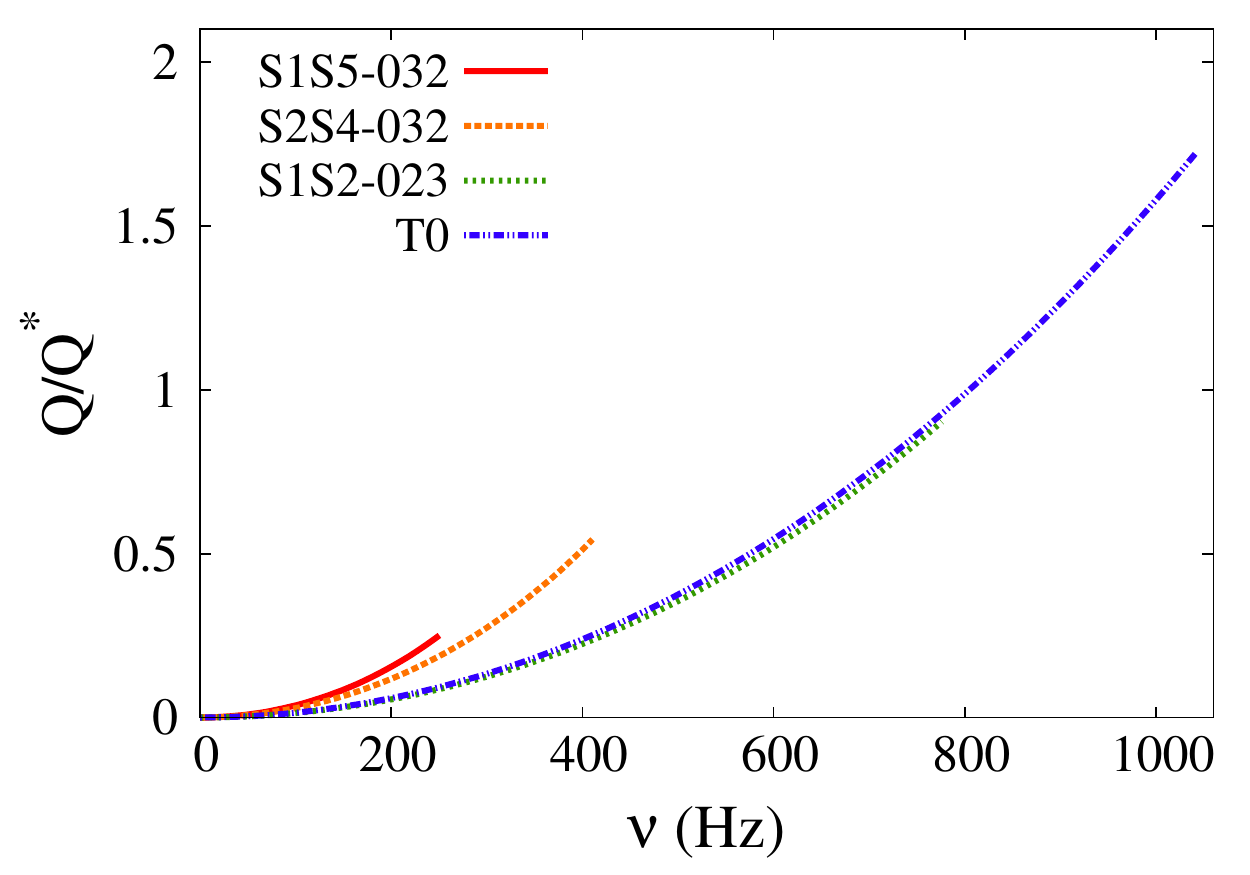}}
\caption{(Color online)
The  mass, equatorial radius, moment of
inertia and quadrupole moment of a star evolving according to the
sequence of profiles \texttt{BS} shown in Fig.~\ref{FIG2}
are plotted  as functions of the rotation rate. For each profile, the
endpoint is the
mass-shedding limit evaluated using the fit (\ref{MSfitDoneva}).
All plots refer to a star with baryonic mass
$M_\tn{b}=1.6~M_\odot$. Moreover $I^* = 10^{45}$ g$\times$cm$^2$ and $Q^* = 10^{44}$
g$\times$cm$^2$.
}
\label{FIG4}
\end{figure*}
%%%%%%%%%%%%%%%%%%%%%%%%%%%%%%%%%%%%%%%%%%%%%%%%%%%%%%%%%%%%%%%%%%
%%%%%%%%%%%%%%%%%%%%%%%%%%%%%%%%%%%%%%%%%%%%%%%%%%%%%%%%%%%%%%%%%%
\subsection{Rotating configurations  \label{secVI}}
%%%%%%%%%%%%%%%%%%%%%%%%%%%%%%%%%%%%%%%%%%%%%%%%%%%%%%%%%%%%%%%%%%
Starting from the non-rotating configurations
belonging to the \texttt{BS} sequence
we can now construct models of rotating proto-neutron stars. 
We choose a value of the baryonic mass and, keeping this value fixed,
we use Hartle's procedure to find how the gravitational mass, 
the circumferential equatorial radius,
the moment of inertia and the quadrupole moment change as functions of the
rotation rate. 
The results are shown in
Fig.~\ref{FIG4}  for a star with $M_\tn{b}=1.6~M_\odot$.
For each profile of the \texttt{BS} sequence, the values of these
quantities are plotted up to the corresponding mass-shedding limit
$\nu_{\tn{ms}}$ evaluated using the fit (\ref{MSfitDoneva}) and
given in the last column of Table~\ref{table1}.
The behaviour of $M, R_{\tn{eq}}, I, Q$ is similar to that of the cold stars
shown in Fig.~\ref{FIG9} (see the Appendix), and all quantities increase with rotation.
However, there is a very interesting difference. The mass shedding
frequency of cold stars, for instance those  considered in
the Appendix, varies in
the range $\sim (800-1350)$ Hz for a star with $M_\tn{b}=1.55~M_\odot$, i.e. by
less than a factor 2 depending on the EoS. 
Conversely for the hot \texttt{BS} sequence it varies
in a much broader range, being $\nu_{\tn{ms}}=256$ Hz for the ``initial''
configuration \texttt{S1S5-032} and $\nu_{\tn{ms}}=1060$ Hz for the
``final'' configuration \texttt{T0} (see Table~\ref{table1}). 
This means that this proto-neutron star cannot enter
the quasi-stationary regime being very rapidly rotating, and that it 
can reach larger rotation rates 
only at subsequent times, as it contracts and cools down.

Rotation rates of newly born proto-neutron stars can be larger
if the baryonic mass is larger because $\nu_{\tn{ms}}$ increases with 
$M_\tn{b}$, as shown in Table~\ref{table2}. In this
case, the behaviour of the relevant quantities is
similar to that shown in Fig.~\ref{FIG4} for $M_\tn{b}= 1.6~M_\odot$.

%%%%%%%%%%%%%%%%%%%%%%%%%%%%%%%%%%%%%%%%%%%%%%%%%%%%%%%%%%%%%%%
\subsubsection{Mass-radius relation
\label{subsecI}}
\begin{figure}[htbp]
\centering
\subfigure{\includegraphics[width = 8cm]{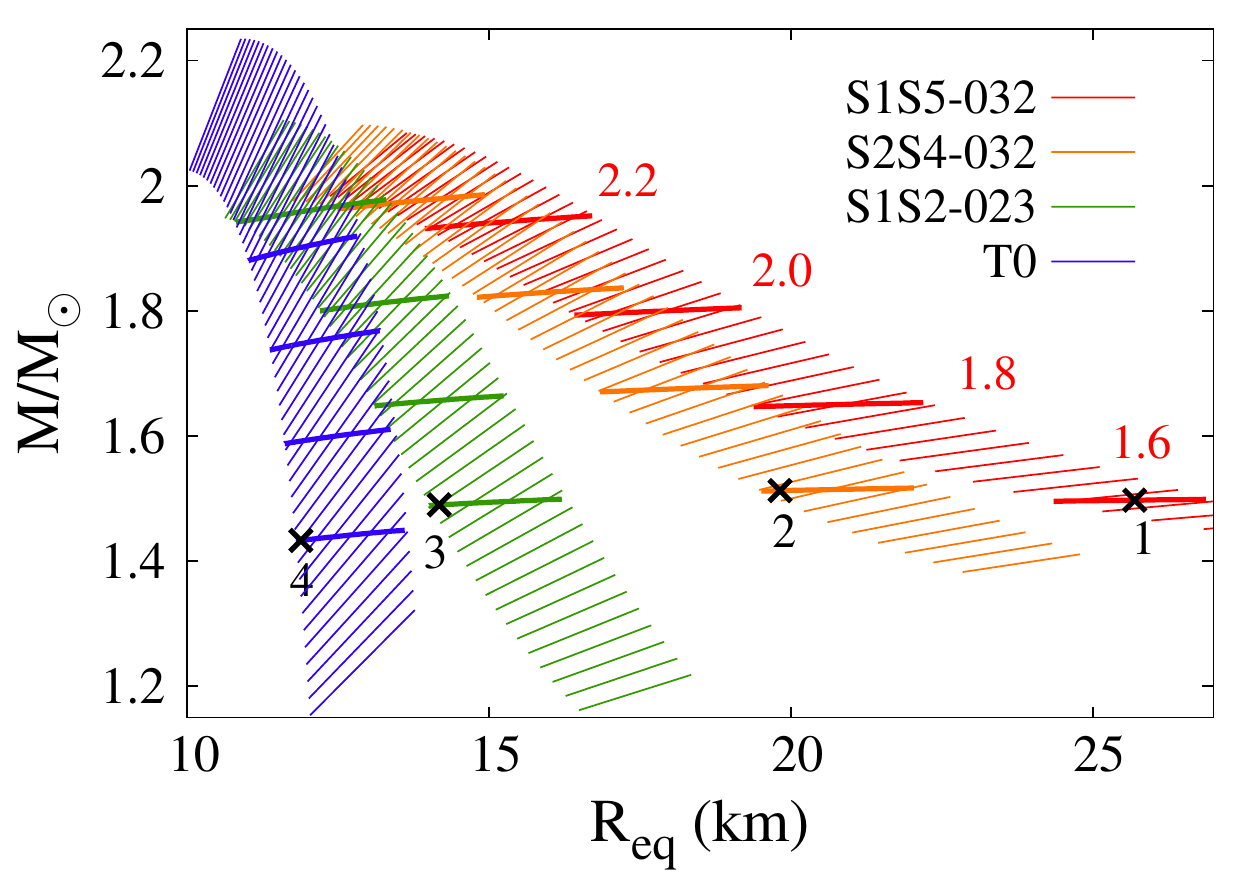}}

\caption{(Color online) Mass-radius diagram for four EoS of the \texttt{BS} set. We
have considered angular frequencies from 0 Hz up to the mass-shedding
limit. Thick lines represent sequences at fixed baryonic masses 
 $M_{\tn{b}}=[1.6, 1.8,2.0,2.2]M_{\odot}$ (see text).}

\label{FIG5}
\end{figure}

%%%%%%%%%%%%%%%%%%%%%%%%%%%%%%%%%%%%%%%%%%%%%%%%%%%%%%%%%%%%%%%
Using the \texttt{BS} profiles  we can construct the
mass-radius diagram for rotating proto-neutron stars. In 
Fig.~\ref{FIG5} we show this diagram  
from the  hottest,  youngest configurations (high entropy gradient
\texttt{S1S5}-\texttt{032}) to the coldest, oldest one (\texttt{T0}). 
%FFFFFFFFFFFFFFFFFFFFFFFFFFFFFFFFFFFFFFFFFFFFFFFFFFFFFFFFFFFFFFFFFF
%FFFFFFFFFFFFFFFFFFFFFFFFFFFFFFFFFFFFFFFFFFFFFFFFFFFFFFFFFFFFFFFFFF
For each profile, the thin, inclined lines  indicate sequences at
fixed central energy density. Each of these lines ends at the
mass-shedding frequency evaluated with the fit (\ref{MSfitDoneva}).
The bold, nearly horizontal lines are the sequences at fixed baryonic
mass, with values $M_\tn{b}=1.6, 1.8, 2.0, 2.2$ M$_\odot$.

As an example, let us consider the $M_\tn{b}=1.6 M_\odot$ star with the
\texttt{S1S5-032} profile. This initial PNS configuration is a point
on the lowest red bold line in Fig.~\ref{FIG5}, and its location on this
line depends on the PNS angular velocity. We indicate the initial
configuration with a `1' on the figure. As the star cools down, it
is constrained to evolve keeping the baryonic mass fixed; therefore it
will jump  toward the left of the figure, reaching the point `2' of the
lowest bold line of the \texttt{S2S4-032} profile (in this example we
keep the angular momentum fixed as well). Then it will move to
the point `3' of the \texttt{S1S2-023} profile and finally to the point
`4' of the cold, final  \texttt{T0} profile, always following the lowest bold lines.
Thus, Fig.~\ref{FIG5} can be understood as a ``temporal'' diagram.
At the end of the evolution  the gravitational mass  is smaller than
that at the beginning.
Since the maximum baryonic mass of the \texttt{S1S5-032} profile
is smaller than that of the final \texttt{T0} profile, 
even if the initial mass of the star is close to the maximum mass
allowed by the  \texttt{S1S5-032} profile, at the end of the
quasi-stationary evolution it will have a gravitational mass smaller
than the maximum mass allowed by the \texttt{T0} profile.

We expect this feature  to be independent of the particular evolutionary
model we consider.

Thus, the maximum mass observed for an isolated  neutron star is not
constrained by the actual (cold) profile, but by the profile it had at
the early stage of its life.

An interesting consequence of this result is the following.
When the gravitational mass of a  NS is known from astrophysical observations,
the EoS which predict a maximum mass larger than this value
are considered compatible with the observation, while the others are
ruled out \cite{Lattimer:2006xb}.  
However, our results show that this ``compatibility test'' 
is a necessary, but not sufficient condition:
indeed, the final NS mass is the result of the evolution  of a hot PNS,
and the cold EoS which should be ``admitted''  are those which result from
an evolutionary sequence compatible with the observed mass.

%%%%%%%%%%%%%%%%%%%%%%%%%%%%%%%%%%%%%%%%%%%%%%%%%%%%%%%%%%%%%%%%%
\subsubsection{Spin frequency change during the quasi-stationary evolution
\label{subsecII}}
%%%%%%%%%%%%%%%%%%%%%%%%%%%%%%%%%%%%%%%%%%%%%%%%%%%%%%%%%%%%%%%%%
We shall now consider a PNS  with  assigned baryonic mass and 
initial  rotation rate $\nu_{\tn{in}}$. 
This star has the \texttt{BS} profile
\texttt{S1S5-032}, which we consider as a model for a PNS at
approximately $0.5$ seconds from bounce. 
We want to compute how the spin frequency changes as the star evolves
along the \texttt{BS} sequence, assuming either that  angular momentum is conserved,
or that a fraction of the angular momentum is lost due to neutrino
emission. To model this second case, we use the empirical formula given
in \cite{2004IAUS..218..3J} (see also references therein) :
\begin{equation}
   \frac{J_{\tn{fin}}}{J_{\tn{in}}} = \left( \frac{M_{\tn{fin}}}{M_{\tn{in}}} \right)^{2.55}~,
\label{janka}
\end{equation}
where $J_{\tn{in}}, M_{\tn{in}}$ and $J_{\tn{fin}}, M_{\tn{fin}}$ are the initial and final
values of the angular momentum and of the star gravitational mass.
In our case, the final mass is that corresponding to the profile
\texttt{T0}. 
The exponent of the mass ratio  in Eq.~(\ref{janka}) takes into account
the shape and structure of the star as well as neutrino efficiency in
removing angular momentum. The value 2.55 results from the most extreme
case, when the angular momentum loss due to neutrinos is maximum.

We shall make the calculations for three values of the baryonic mass
(kept fixed along the sequence)
$M_\tn{b}=1.6, 1.8, 2.0$ M$_\odot$, 
 and for four values of
the initial rotation rate, $\nu_{\tn{in}} = (0.25,0.5, 0.75, 1)
\nu_{\tn{ms}}$.

In Table~\ref{table3} we show the results.

%TTTTTTTTTTTTTTTTTTTTTTTTTTTTTTTTTTTTTTTTTTTTTTTTTTTTTTTTTTTTTTTT
\begin{table}[htbp]
\centering
\begin{tabular}{ccc|ccc|ccc}
\hline
\hline
\multicolumn{9}{c}{$M_\tn{b} = 1.6 M_{\odot}$ $(M_{\tn{in}}=1.50 M_{\odot},\  
M_{\tn{fin}}=1.43 M_{\odot})$}\\
\hline
\multicolumn{3}{c}{}&\multicolumn{3}{|c|}{$J$ conserved}&
\multicolumn{3}{c}{neutrino losses}\\
\hline
$\bar{\nu}_{\tn{in}}$&  $\nu_{\tn{in}}$   & $J_{\tn{in}}$ 
& $\bar{\nu}_{\tn{fin}}$  &  $\nu_{\tn{fin}}$  & $J_{\tn{fin}}$
& $\bar{\nu}_{\tn{fin}}$  &  $\nu_{\tn{fin}}$  &
$J_{\tn{fin}}$\\
\hline
 0.25 & 64.0 & 0.184 & 0.079 & 83.8 & 0.184 & 0.071 & 74.9 & 0.164 \\ \hline
 0.50 & 128  & 0.371 & 0.159 & 169  & 0.371 & 0.142 & 151  & 0.332 \\ \hline
 0.75 & 192  & 0.566 & 0.241 & 255  & 0.566 & 0.216 & 229  & 0.507 \\ \hline
 1.00 & 256  & 0.773 & 0.325 & 345  & 0.773 & 0.292 & 310  & 0.691 \\ \hline
\hline
\multicolumn{9}{c}{$M_\tn{b} = 1.8 M_{\odot}$ $(M_{\tn{in}}=1.65 M_{\odot},\ 
M_{\tn{fin}}=1.59 M_{\odot})$}\\
\hline
 0.25 & 114  & 0.321 & 0.112 & 129  & 0.321 & 0.102 & 118  & 0.293 \\ \hline
 0.50 & 228  & 0.651 & 0.226 & 260  & 0.651 & 0.206 & 237  & 0.593 \\ \hline
 0.75 & 342  & 0.997 & 0.341 & 392  & 0.997 & 0.312 & 359  & 0.908 \\ \hline
 1.00 & 456  & 1.37  & 0.461 & 530  & 1.37  & 0.423 & 486  & 1.25  \\ \hline
\hline
\multicolumn{9}{c}{$M_\tn{b} = 2.0 M_{\odot}$ $(M_{\tn{in}}=1.79 M_{\odot},\ 
M_{\tn{fin}}=1.74 M_{\odot})$}\\
\hline
 0.25 & 165  & 0.478 & 0.140 & 173  & 0.478 & 0.129 & 160  & 0.440 \\ \hline
 0.50 & 329  & 0.968 & 0.281 & 348  & 0.968 & 0.259 & 321  & 0.892 \\ \hline
 0.75 & 494  & 1.49  & 0.423 & 524  & 1.49  & 0.392 & 486  & 1.37  \\ \hline
 1.00 & 658  & 2.07  & 0.569 & 706  & 2.07  & 0.530 & 657  & 1.91  \\ \hline
\hline
\end{tabular}
\caption{
Values of the spin frequency that a star, with a fixed baryonic mass and
an initial rotation rate $\nu_{\tn{in}}$,
reaches at the end of the quasi-stationary evolution,
assuming that the angular momentum   is
conserved  or that it changes due to neutrino emission.
$J$ is given in km$^2$, frequencies are in Hz.
The initial values refer to a proto-neutron star  modeled with the
profile \texttt{S1S5-032}, the final ones refer to the cold star
with profile \texttt{T0}. Barred frequencies are normalized to the 
mass-shedding frequency of the initial and final profiles.
}
\label{table3}
\end{table}
%TTTTTTTTTTTTTTTTTTTTTTTTTTTTTTTTTTTTTTTTTTTTTTTTTTTTTTTTTTTTTTTTTTTTTTTTTT

As an example, let us consider a PNS with $M_\tn{b}= 1.6~M_\odot$ which, a
few hundreds of ms after bounce, has the profile
\texttt{S1S5}-\texttt{032}. The mass-shedding frequency for this profile
is $\nu_{\tn{ms}}=256$ Hz, and we assume that the star has initial 
rotation rate $\nu_{\tn{in}}=0.25 \nu_{\tn{ms}} = 64$ Hz.
The angular momentum of this configuration is
$J= 0.184$ km$^2$, the gravitational mass is $M_{\tn{in}}=1.50 M_\odot$ (see
Table~\ref{table1}). 

Using Hartle's procedure we  evaluate the rotation rate that this
star reaches at the end of the quasi-stationary evolution, when it has
the profile \texttt{T0}, i.e. zero temperature,
smaller radius and moment of inertia, same baryonic mass
and smaller gravitational mass due to neutrino emission ($M_{\tn{fin}}=1.43 M_\odot$). 
If we impose that the angular momentum is conserved,
from Table~\ref{table3}  we see that the final  rotation rate will be
$\nu_{\tn{fin}}=83.8$ Hz,
corresponding to $0.079 $ times the mass-shedding limit for the
\texttt{T0} profile, which is $\nu_{\tn{ms}}=1060$ Hz. 
If, conversely, we assume that the angular momentum changes according to
Eq.~(\ref{janka}), the final spin frequency will be lower,
$\nu_{\tn{fin}}=74.9$ Hz, corresponding to $\sim 0.071 $ of the
mass-shedding limit.

Thus, if the star is born with a
rotation rate which is about one fourth of the mass-shedding limit, its rotation
speed will remain slow even at the end, when the mass shedding limit
would, in principle, allow for larger rates.

If the star starts with a larger rotation rate, say at the mass-shedding
limit, at the end of the evolution its frequency will be larger, 
$\nu_{\tn{fin}}= 345$ Hz, or $\nu_{\tn{fin}}= 310$ Hz accounting for neutrino
losses, but always smaller than the final mass-shedding limit.
These considerations hold also for stars with larger baryonic mass.

Last but not least, the sequences we have considered show that the ratio
$\nu/\nu_{\tn{ms}}$, even if large at the beginning of the
evolution (large entropy gradient), constantly decreases as one goes
along the cooling of the star. Thus, mature  NS evolved in isolation
cannot rotate too rapidly, even if they are born from a
PNS rotating at the mass-shedding limit
\footnote{We would like to remind that from a naive point of view, a
spherical Newtonian fluid ball has
angular momentum $I\Omega \propto MR^2\Omega$ and mass-shedding limit
$\propto M^{1/2}R^{-3/2}$. A factor  2 of decrease in radius would lead
to an increase of a factor 4 in $\Omega$ and of a factor  $\sqrt{8} \sim 3$ in the
mass-shedding limit,
thus allowing the final configuration to be close to the mass-shedding limit if
the initial one was.}.

%FFFFFFFFFFFFFFFFFFFFFFFFFFFFFFFFFFFFFFFFFFFFFFFFFFFFFFFFFFFFFFFF
\begin{figure}[ht]
\centering
\includegraphics[width=8.5cm]{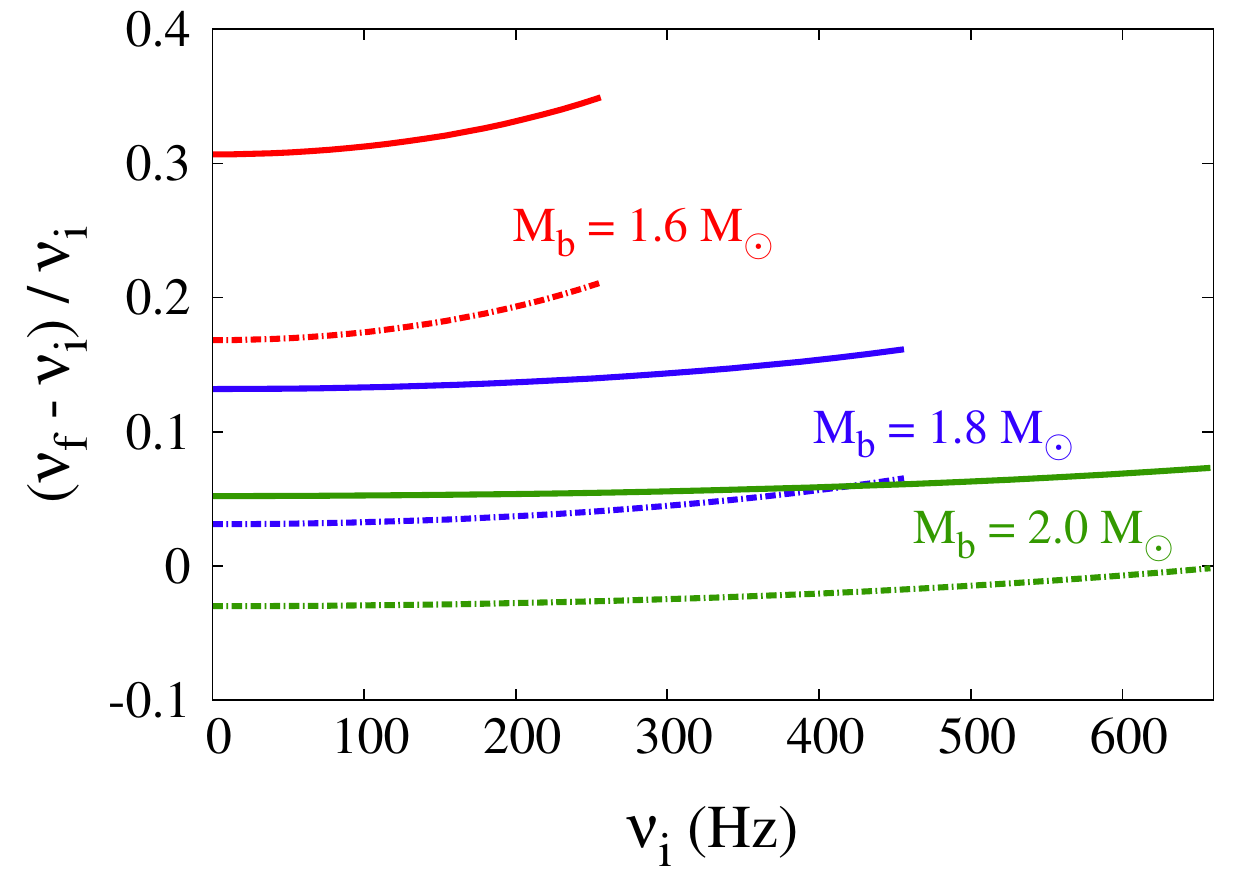}
\caption{(Color online) We show the relative difference  between the initial
and final spin frequency, as a function of  the initial spin rate,
spanning the range  from zero to the mass-shedding limit.
The results are shown for three values of the baryonic mass;
continuous lines refer to an evolution which assumes  angular momentum 
conservation, dashed lines  neutrino emission.}
\label{FIG6}
\end{figure}
%FFFFFFFFFFFFFFFFFFFFFFFFFFFFFFFFFFFFFFFFFFFFFFFFFFFFFFFFFFFFFFFF
We have further explored the entire range of initial spin rate,
from zero to the mass-shedding limit, computing the corresponding
final spin frequency, for different values of the    
baryonic mass. 
In Fig.~\ref{FIG6} we plot the relative difference
$(\nu_{\tn{fin}}-\nu_{\tn{in}})/\nu_{\tn{in}}$, for the considered values of the  
baryonic mass, as a function of $\nu_{\tn{in}}$.
We see that for the lightest star the change reaches $\sim 35\%$ without 
including neutrino emission, and $\sim 25\%$ with neutrino emission, whereas for the 
heaviest  star the percentual change does not exceed $\sim 5\%$. In this case 
it should be noted that the change may be negative for small values of 
$\nu_{\tn{in}}$, showing that stellar contraction due to the cooling,
and neutrino emission could be competitive in some regime. 
This last result is EoS dependent, but still provides a proof-of-concept.

It should also be reminded that, as shown in section~\ref{secIII} and
in the Appendix,  values of frequency too
close to the mass-shedding limit are associated to larger errors on the
various quantities, which for the
moment of inertia can reach  $\sim 11\%$; however the errors rapidly decrease for
lower rotation rates.

The early evolution of a PNS has also been studied in
\cite{Dexheimer:2008ax}, using a mean field EoS 
(based on a chiral $SU(3)$ description of baryons) and  assuming isentropic 
or isothermal profiles. They compute the maximum mass and the
mass-shedding frequency for different values of entropy and temperature, 
assuming constant baryonic mass and constant angular momentum.
This is different from what we do, since they do not compute the
evolution of the gravitational mass and of the rotation rate.  

There is another  process which may subtract angular momentum to the PNS,
if this is born with a certain degree of asymmetry: gravitational wave
(GW) emission.
If the  PNS has a finite ellipticity $\epsilon$, such waves are emitted
at twice the rotation frequency.
However, the spin-down induced by GW during the quasi-stationary
evolution is negligible
with respect to the spin-up due to contraction. A simple estimate shows that
$\frac{\delta\Omega}{\Omega}\sim -\frac{L_{GW}\delta t}{I\Omega^2}-\frac{\delta
  I}{2I}$ (where $L_{GW}=\frac{32G}{5c^5}\Omega^6\epsilon^2I^2$); assuming
for instance $\epsilon\sim10^{-6}$, $I\sim10^{45}$ g$\times$ cm$^2$,
$\Omega\sim 1$ kHz, and considering that in a timescale $\delta t\sim20 s$ the
moment of inertia is reduced by a factor  $\sim 2$,
the GW contribution to the spin-down is
$\sim10^9$ smaller than that due to contraction.

During the first few hundreds of milliseconds after bounce, the GW
signal is expected to be dominated  by processes associated to neutrino emission,
magnetorotational processes, SASI, dynamical instabilities
etc. \cite{Scheid:10,Logue:2012zw,Ott:2012mr}. 
However, we have just shown that when the evolution becomes quasi-stationary
the signal could be dominated by the spin-up due to contraction.  
Thus in this phase, which would last for about a minute,
the signal would be ``chirp-like'', 
and this feature may be searched in the post-processing of the data associated
to a GW signal from a supernova with appropriate data analysis techniques.

Finally, we would like to mention  that in our study we did not include the effect
of differential rotation, which may be present in the very early stages
of the PNS evolution. This effect will be studied in a separate paper. 

%%%%%%%%%%%%%%%%%%%%%%%%%%%%%%%%%%%%%%%%%%
%%%%%%%%%%%%%%%%%%%%%%%%%%%%%%%%%%%%%%%%%%

\section{I-Love-Q relations for hot and young neutrons stars
\label{secVII}}

As discussed in the introduction, it has recently been found that the moment of 
inertia ($I$), the tidal deformability ($\lambda$)  and the spin-induced quadrupole 
moment ($Q$) of isolated and binary neutron stars, are related by universal 
relations which are independent of the star internal composition 
\cite{Yagi:2013bca,Yagi:2013awa,Maselli:2013mva,Doneva:2013rha,Haskell:2013vha} 
 (further universal relations have been studied in
\cite{Baubock:2013gna,Stein:2014wpa,Pappas:2013naa,Yagi:2014bxa}).

Until now, such I-Love-Q relations have been tested
only for cold EoS.  In \cite{Yagi:2013bca,Yagi:2013awa}
two finite temperature EoS are also  employed, LS220 and
Shen.  However, these EoS are treated as barotropic, assuming a
uniform temperature of $T\simeq10^{9}$ K and no neutrinos: they
describe the star more than one minute after the bounce. 

%TTTTTTTTTTTTTTTTTTTTTTTTTTTTTTTTTTTTTTTTTTTTTTTTTTTTTTTTTTTTTTT
\begin{table}[ht]
  \centering
  \begin{tabular}{ccccccc}
   \hline
    \hline
     $y$ & $x$ & $a$ & $b$ & $c$ & $d$ & $e$\\  
     \hline
 $\bar{I}$ & $\bar{\lambda}$ & $1.47$ & $0.0817$ & $0.0149$ & $0.000287$ & $-3.64\cdot10^{-5}$\\
 $\bar{I}$ & $\bar{Q}$ & $1.35$ & $0.697$ & $-0.143$ & $0.0994$ & $-1.24\cdot10^{-2}$\\
  $\bar{Q}$ & $\bar{\lambda}$ & $0.194$ & $0.0936$ & $0.0474$ & $-0.00421$ & $1.23\cdot10^{-4}$\\
  \hline
  \hline
  \end{tabular}  
  \caption{Best-fit coefficients of Eq.~(\ref{YY}) for the I-Love-Q relations
   \cite{Yagi:2013bca}.}\label{table4}
  \end{table}
%TTTTTTTTTTTTTTTTTTTTTTTTTTTTTTTTTTTTTTTTTTTTTTTTTTTTTTTTTTTTTTT
In this section, we will assess the range of validity of the universal
relations for newly-born stars, in which an entropy gradient between the
core and the envelope is still present. Indeed, we will compare the
numerical data obtained computing the $I$-$\lambda$-$Q$ trio for NS
described by the  profiles \texttt{GM3NQ} (see section~\ref{secIV}),
with the relations originally found in \cite{Yagi:2013bca,Yagi:2013awa}.
We remind that the \texttt{GM3NQ} profiles describe the evolution of a
hot proto-neutron star with baryonic mass $M_\tn{b}=1.6M_{\odot}$ during
the first minute of life after the gravitational collapse.

The fits have the following functional form:
\begin{equation}\label{YY}
\ln y= a+b\ln x+c(\ln x)^2+d(\ln x)^3+e(\ln x)^4
\end{equation}
where $(a,b,c,d,e)$ are fitting coefficients listed in
Table~\ref{table4}.
Eq.~(\ref{YY}) is defined in terms of the normalized variables
$(\bar{I},\bar{Q},\bar{\lambda})$, where $\bar{I}=I/M^3$,
$\bar{\lambda}=\lambda/M^5$ and $\bar{Q}=Q (M/J^2)$,
being $M$ the neutron star gravitational mass and $J$ its
angular momentum.
%TTTTTTTTTTTTTTTTTTTTTTTTTTTTTTTTTTTTTTTTTTTTTTTTTTTTTTTTTTTTTTT
\begin{table}[ht]
  \centering
\begin{tabular}{ccccccc}
\hline
    \hline
     t(s) &$M(M_{\odot})$  
     & $R$(km) & $\bar{Q}$ & $\bar{I}$ & $\bar{\lambda}$ \\  
     \hline
  0.2 & 1.58  & 34.39 & 18.86 & 31.32 &   12999.60      \\
  0.3 & 1.57  & 28.98 & 15.23 & 26.83 & 7374.64        \\
 0.5  & 1.56  & 23.79 & 12.04 & 22.59 &    3630.51     \\
 1  & 1.55  & 19.38 & 9.40  & 18.70 & 1914.93        \\
 2  & 1.53  & 15.76 & 7.39  & 15.58 &  945.85       \\
 5  & 1.50  & 13.61 & 6.20  & 13.56 &  564.98       \\
 20 & 1.47 & 12.91 & 5.80  & 12.83 & 458.91        \\
 \hline
 \hline
\end{tabular}
\caption{Parameters of the NS models with baryonic mass 
$M_{\tn{b}}=1.6M_{\odot}$ built with the \texttt{GM3NQ} EoS. 
In the first column we show the time after the star  birth. 
In the remaining columns we list the gravitational mass 
$M$, the radius at spherical equilibrium, the quadrupole 
moment $\bar{Q}$, the moment of inertia $\bar{I}$, and the 
tidal deformability $\bar{\lambda}$.}
\label{table5}
\end{table}
%TTTTTTTTTTTTTTTTTTTTTTTTTTTTTTTTTTTTTTTTTTTTTTTTTTTTTTTTTTTTTTT
We have computed $\bar{I}$, $\bar{\lambda}$ and $\bar{Q}$ for the
\texttt{GM3NQ} proto-neutron star at different stages of evolution, i.e.
 at $t=(0.2,0.3,0.5,1,2,5,20)$ seconds after
birth. The corresponding values are shown in Table~\ref{table5}.

In order to determine the accuracy within which the I-Love-Q relations
would describe the features of a newly-born neutron star, 
we have computed the relative
errors
$\Delta I /\bar{I}_{\tn{fit}}
=\vert\bar{I}-\bar{I}_{\tn{fit}}\vert/\bar{I}_{\tn{fit}}$ and
$\Delta Q /\bar{Q}_{\tn{fit}}
=\vert\bar{Q}-\bar{Q}_{\tn{fit}}\vert/\bar{Q}_{\tn{fit}}$,
between our numerical data, and those
obtained by using the fit (\ref{YY}).
%
%FFFFFFFFFFFFFFFFFFFFFFFFFFFFFFFFFFFFFFFFFFFFFFFFFFFFFFFFFFFFFF
\begin{figure}[ht]
\centering
\includegraphics[width=8.5cm]{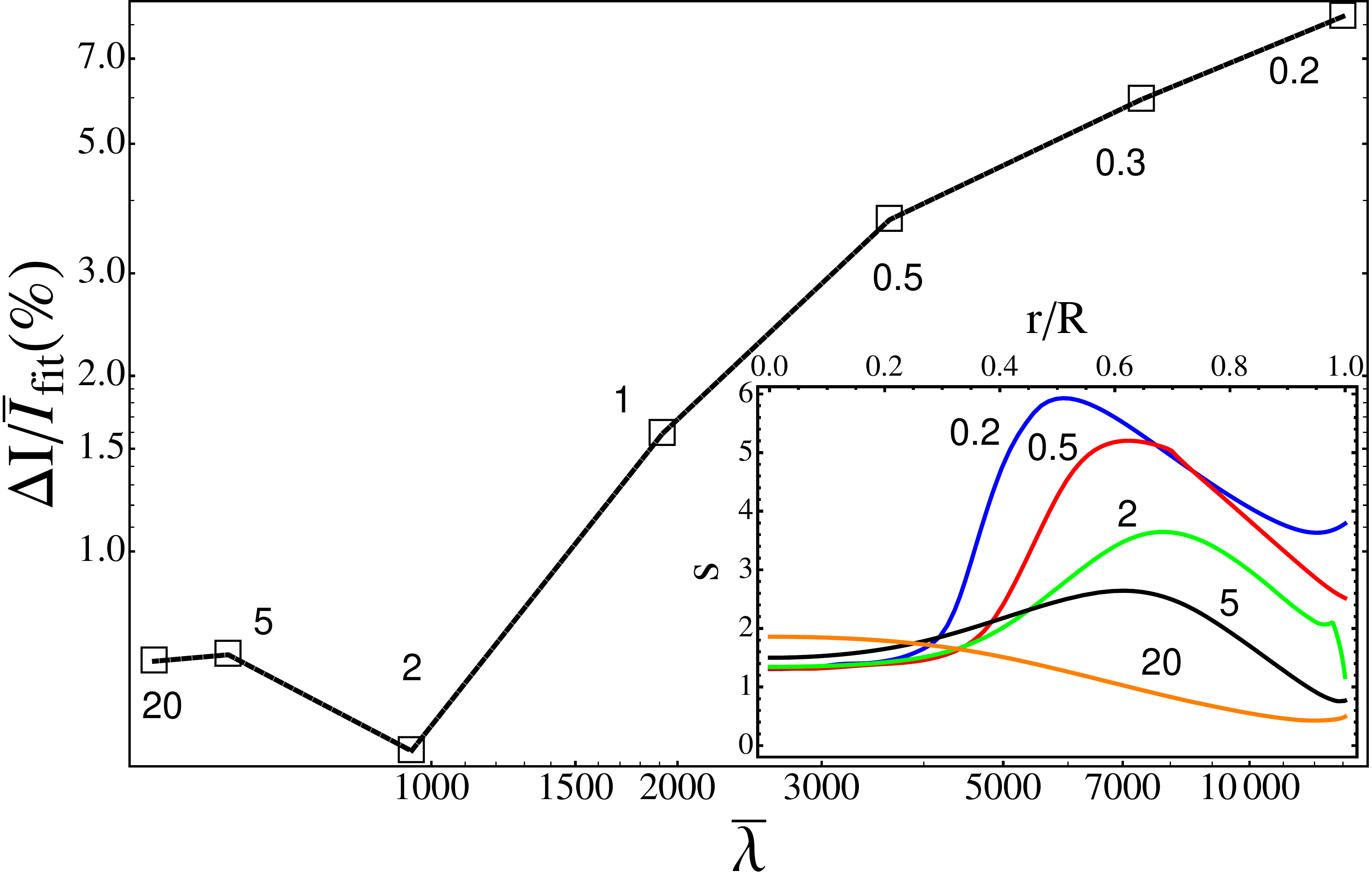}
\includegraphics[width=8.5cm]{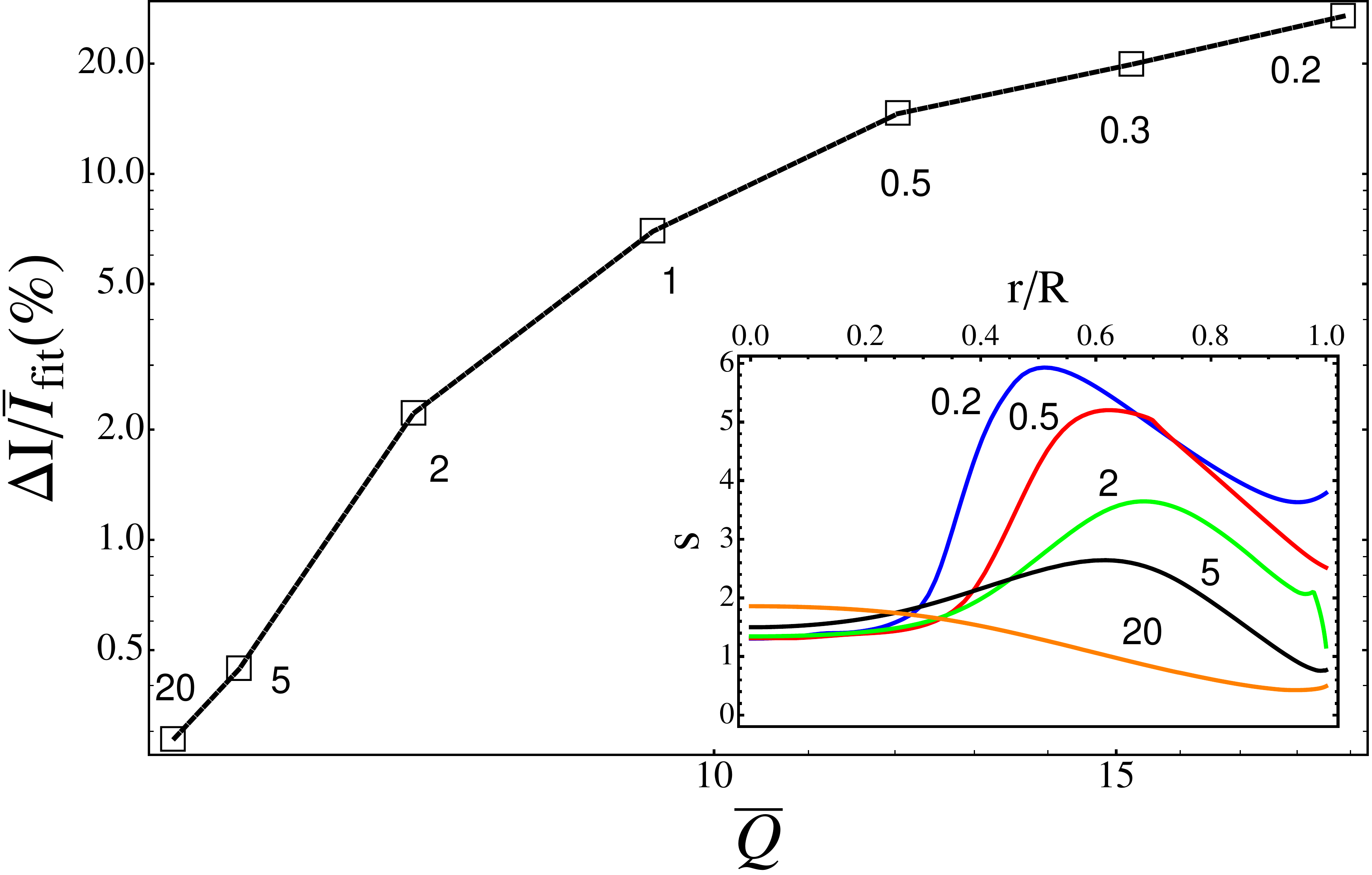}
\includegraphics[width=8.5cm]{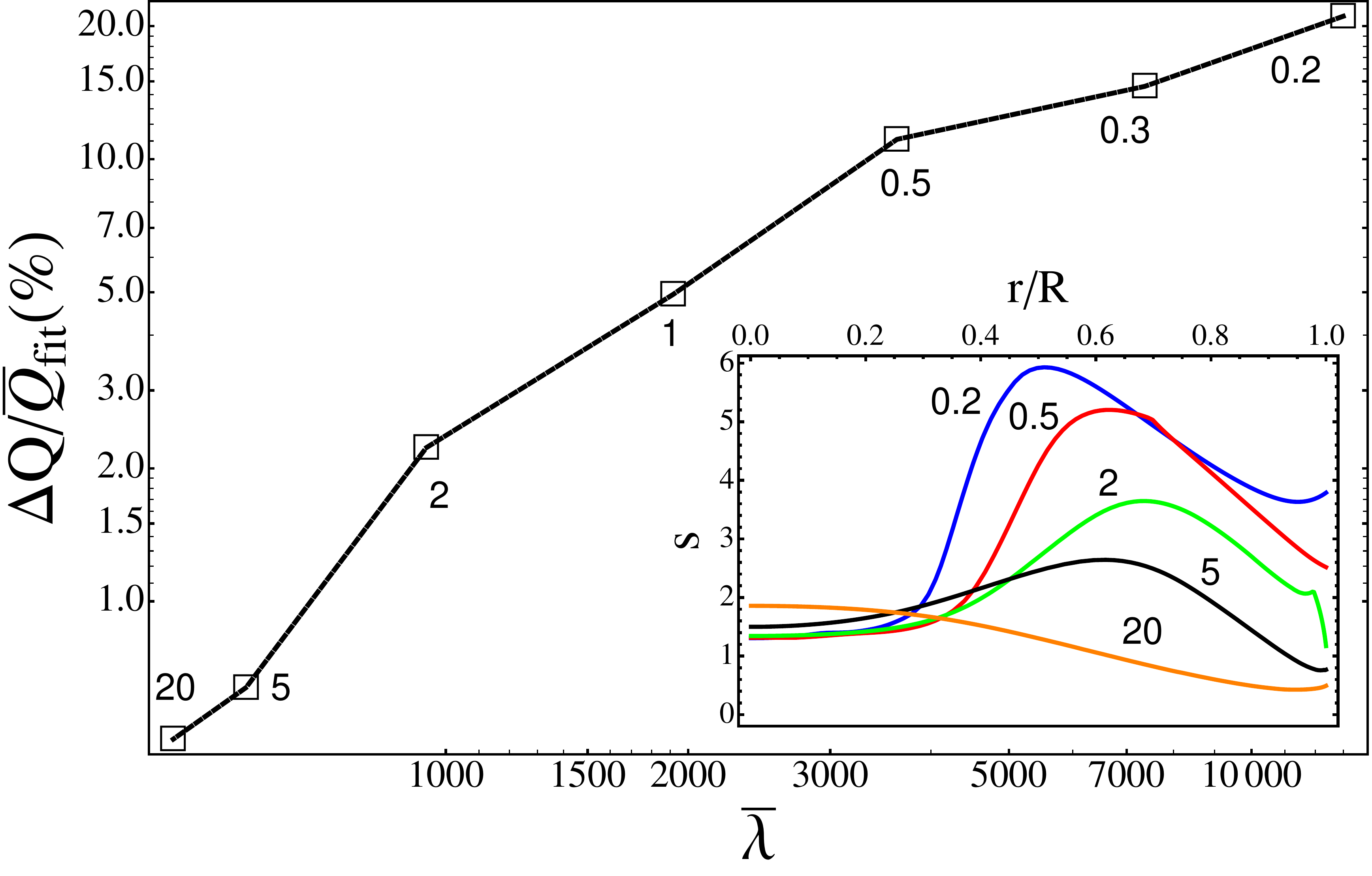}
\caption{(Color online) We show the relative difference
$\vert \bar{I}-\bar{I}_{\tn{fit}}\vert/\bar{I}_{\tn{fit}}$ and
$\vert \bar{Q}-\bar{Q}_{\tn{fit}}\vert/\bar{Q}_{\tn{fit}}$
obtained testing the universal relations (\ref{YY})
against the numerical data computed using the
\texttt{GM3NQ} equations of state, at different times (in seconds)
after the birth of the proto-neutron star. In the inset we also show for
some of the considered configurations, the entropy per
baryon as function of the star radius.}
\label{FIG7}
\end{figure}
%FFFFFFFFFFFFFFFFFFFFFFFFFFFFFFFFFFFFFFFFFFFFFFFFFFFFFFFFFFFFFF

The results are summarized in
the three panels of Fig.~\ref{FIG7}, which
clearly show  that the I-Love-Q relations lose 
their validity in the very early stages after the star birth, 
with discrepancies between the 
analytic fit and the numerical values which can be as high as $\sim 30\%$ at
$t=0.2$ s, for the $\bar{I}$-$\bar{Q}$ pair (middle panel
in Fig.~\ref{FIG7}).

However, the relative errors rapidly decrease as time increases: 
after $1$ second, the relative difference between  our moment of inertia
and the fit $\bar{I}$-$\bar{\lambda}$ is $1.6~\%$, 
that with respect to the fit $\bar{I}$-$\bar{Q}$ is $6.7~\%$,  and the relative
difference between our quadrupole moment with respect to the fit
$\bar{Q}$-$\bar{\lambda}$ is
$4.8~\%$. After $2$ seconds these errors reduce to $< 1~\%$, $2.2~\%$,
and $2.0~\%$, respectively.
It it interesting to analyze these results 
in terms of the entropy gradient inside the star at different times.
In each plot of Fig.~\ref{FIG5} we have included the plot of
the entropy per baryon, as a function of the distance from the center 
of the star (normalized to the radius $R$),  for some of the 
considered configurations. This shows that   the largest 
relative errors correspond to the higher entropy gradients inside the star 
(blue and red curves), which develop after its birth and smooth 
down during the following evolution.
Thus, the time interval during which the I-Love-Q relations are
accurate depends on how fast 
the star reaches the quiet state of a neutron star.

\begin{figure}[ht]
\centering
\includegraphics[width= 7.5cm]{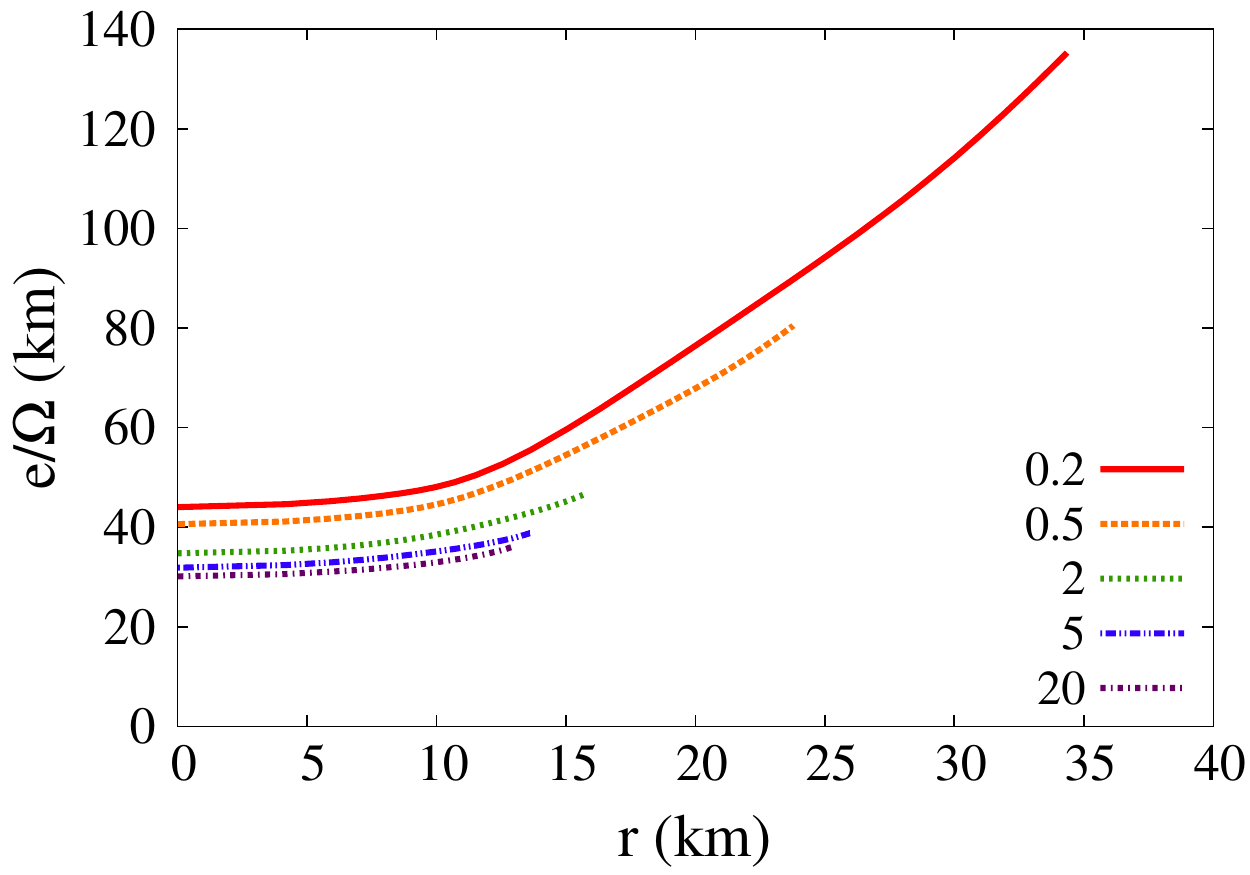}
\caption{(Color online) The ellipticity  of the isodensity
  contours, normalized to the rotation rate, is plotted as a function of
the radial distance at different times of the PNS
  evolution. This plot refers to  the \texttt{GM3NQ} quasi-stationary sequence with
  $M_\tn{b}=1.6\,M_\odot$.}
\label{FIG8}
\end{figure}

It has recently been suggested \cite{Yagi:2014qua} that the I-Love-Q relations
become less accurate when the ellipticity of the isodensity contours has a
large gradient inside the star. In Fig. \ref{FIG8} we show the behaviour of the
ellipticity $e$ of the isodensity contours (see Eq.~(25c) of
\cite{Hartle:1968si}), normalized to  the star rotation rate, for the
\texttt{GM3NQ} quasi-stationary sequence with $M_\tn{b}=1.6\,M_\odot$ considered
in this Section. We see that in the first second after bounce, when the I-Love-Q
relations are violated, the ellipticity exhibits significant variations throughout the
star ($\gtrsim200\%$). At later times, when the entropy profiles smooth out and
the I-Love-Q relations are satisfied, the ellipticity profiles are nearly
constant. This supports the suggestion of \cite{Yagi:2014qua} that the validity
of the I-Love-Q relations is associated with the self-similar isodensity
condition.

%%%%%%%%%%%%%%%%%%%%%%%%%%%%%%%%%%%%%%%%%%%%%%%%%%%%%%%%%%%%%%%%%%
%%%%%%%%%%%%%%%%%%%%%%%%%%%%%%%%%%%%%%%%%%%%%%%%%%%%%%%%%%%%%%%%%%

%%%%%%%%%%%%%%%%%%%%%%%%%%%%%%%%%%%%%%%%%%
\section{Concluding Remarks}
%%%%%%%%%%%%%%%%%%%%%%%%%%%%%%%%%%%%%%%%%%

In this paper we have studied how rotation affects the quasi-stationary
evolution of newly born proto-neutron stars.  
We have used Hartle's perturbative approach at third order in the
angular velocity, which we
have extended to describe warm stars, with non-barotropic EoS. 
Hartle's equations have been integrated
to show how the rotation rate of a PNS with fixed baryonic mass, changes 
during the evolution, taking into account neutrino losses in a heuristic way.

To model the PNS we have used an EoS 
obtained within the  nuclear many-body theory
extended to finite temperature \cite{book,Burgio10,bloch,Burgio11}, and a sequence of 
entropy and energy density profiles which mimics the 
stellar interior of a PNS. This is supposed to evolve from a few tenths of seconds
after bounce (temperature $\sim 30-40$ MeV ($\sim 5\times10^{11}$ K), 
radius $\sim 30$ km, strong entropy gradients), to 
the following minute(s) during which gradients are smoothed out,
temperature decreases to a few MeV, and radius decreases to 10-15 km. 

The EoS and profiles we have used are constructed having as a reference
the quasi-stationary evolution profiles obtained in \cite{Pons99}
by solving Boltzman's equation and using an EoS
obtained within a finite-temperature,
field-theoretical model solved at the mean field level.
To our knowledge, the stellar models given in \cite{Pons99}
are the only example of the quasi-stationary evolution 
of a non rotating PNS available in the literature 
(we also mention the numerical simulations of \cite{Roberts:2011yw,Roberts:2012zza}, 
which employ EoS very similar to those used in \cite{Pons99}, and find similar results). 
Unfortunately we could not use these models  to study how they
change with rotation, because they are given for fixed values of 
the central energy density,
whereas to generate a rotating star with fixed baryonic mass, we need to
change it.
However, we believe that the results we obtain in this paper with 
the \texttt{BS}  sequence give
interesting indications on the quasi-stationary evolution of a rotating, hot PNS,
provided no instability is present (see discussion in
Section~\ref{secI}),
and show a methodology which could be used with other stellar models of
evolving PNS, when available.

Our main results are the following:
\begin{itemize}
\item
As a PNS cools down, the maximum mass allowed by 
the EoS and by the chosen entropy profiles
increases, and the maximum rotation rate (mass-shedding limit) increases as
well. 
   \item A cold neutron star cannot have the maximum mass allowed by
the final zero temperature EoS, unless the
evolutionary path of the hot PNS from which it has evolved allowed to do so. 
It is commonly believed that an equation of state is 
able to describe a NS interior
if the maximum mass it predicts is larger than the maximum mass
observed in neutron stars. However  our results show that
this should be considered as a necessary, but not
sufficient condition.
This means that a full understanding of NS structure cannot overlook the 
history of these objects.

   \item An isolated neutron stars, even when born rotating 
near the mass-shedding limit, 
cannot have a rotation rate close to the  mass-shedding limit
associated to the cold EoS which characterizes its interior (see
Table~\ref{table3} and Figure~\ref{FIG6}).

\item If an isolated NS is found to be rapidly rotating,
Table~\ref{table3} indicates  that its mass must be high.

   \item I-Love-Q relations are no longer valid for very hot neutron
stars, but become valid all the same a few seconds after the bounce. 
Indeed our results show that such relations strongly depend on the 
entropy profile developing inside the star, and that their universality character 
is recovered as soon as entropy gradients smooth out during the evolution.
\end{itemize}
This work should (and will) be extended in many ways to
include more realistic physics  (e.g. differential rotation, magnetic
fields, dynamical neutrino leakage etc.) and a more accurate description of the PNS
evolution.

\section{Acknowledgments}
We would like to thank Fiorella Burgio, Hans-Joseph Schultze and Jos\'e Pons for
kindly allowing us to use the profiles describing the early evolution of a
proto-neutron star.  We also thank Morgane Fortin and Georgios Pappas for useful
comments and discussions.  This work was partially supported by ``NewCompStar''
(COST Action MP1304).

\appendix
%%%%%%%%%%%%%%%%%%%%%%%%%%%%%%%%%%%%%%%%%%%%%%%%%%%%%%%%%%%%%%%%%%%%%%%%%
\section{}
%%%%%%%%%%%%%%%%%%%%%%%%%%%%%%%%%%%%%%%%%%%%%%%%%%%%%%%%%%%%%%%%%%%%%%%%%
In this appendix we compare  some relevant stellar quantities computed
by
integrating Hartle's equations at third order in $\Omega$ (Hartle's
results to hereafter), with those
computed by using the fully relativistic code RNS (RNS results to
hereafter).
In particular, we compute the mass, radius and moment of inertia.

A few words about the EoS we use for this comparison:
\begin{itemize}
   \item \texttt{A} : EoS obtained in \cite{Phandharipande71a} using many-body
theory, and
named EoS \texttt{A} in \cite{Arnett77}. Matter is composed of neutrons only.
Interactions include a Reid soft core adapted to nuclear matter. The
many-body Schrodinger equation is solved with a variational approach,
applied to the correlation function.
  \item \texttt{AU} : EoS named AV14+UVII in \cite{Wiringa88}, matched to
Negele and Vautherin \cite{Negele73} at low densities.  Based on
variational
methods with correlation operators on a Hamiltonian featuring the Argonne
$v_{14}$ two-nucleon potential with the Urbana VII three-nucleon
potential.
\item \texttt{APR} : EoS obtained in \cite{Akmal98} using many-body theory.
Matter is composed of proton, neutrons, electrons and muons.
Interactions are described by the Argonne $v_{18}$ two-nucleon potential
and the Urbana VII three-nucleon potential, including relativistic
corrections arising from the boost to a frame in which the total
momentum of the interacting pair is non-vanishing.
   \item \texttt{O} : EoS obtained in \cite{Bowers75} and named as \texttt{O} in
\cite{Arnett77}. Matter is composed of neutrons, protons and hyperons.
Interactions are non-perturbative, it is a
phenomenological approximation to relativistic meson exchange. The
many-body theory is based on relativistic finite density Green's
functions.
   \item \texttt{g240} : Obtained in \cite{Glendenning:1986zz,g240}.
Matter composition includes leptons and the complete octet of baryons
(nucleons, $\Sigma^{0,\pm}, \Lambda^0$ and $\Xi^{\pm}$). Hadron dynamics
is described in terms of exchange of one scalar and two vector mesons.
The EoS is obtained within the mean field approximation.
\end{itemize}
%TTTTTTTTTTTTTTTTTTTTTTTTTTTTTTTTTTTTTTTTTTTTTTTTTTTTTTTTTTTTTTTTTTT
\begin{table}[h]
\centering
\begin{tabular}{cccccccc}
\hline
\hline
  &    \texttt{g240} & \texttt{APR}  & \texttt{A} && \texttt{O} &
\texttt{APR} & \texttt{AU}               \\
\hline$M_\tn{b}$ &  & $1.55 M_{\odot}$ &&  &
   & $2.2 M_{\odot}$     &   \\
\hline$M/M_{\odot}$  & 1.41 & 1.39 & 1.36 && 1.91 & 1.87 & 1.85  \\
\hline$R$ (km)       & 12.84 & 11.33 & 9.61 && 12.74 & 11.05 & 10.25  \\
\hline
$\bar\rho$ (km$^{-1}$)  & 0.031 & 0.038 & 0.048 && 0.037 & 0.045 &
0.050\\
\hline 
$M/R$  & 0.16 & 0.18 & 0.21 && 0.22  & 0.25  & 0.27
\\
\hline
\hline
\end{tabular}
\caption{Parameters of the non rotating stellar configurations which we
use to
compare the results of the perturbative approach to those of fully
relativistic calculations.  $\bar\rho=\sqrt{M/R^3}$.
}
\label{table6}
\end{table}
%TTTTTTTTTTTTTTTTTTTTTTTTTTTTTTTTTTTTTTTTTTTTTTTTTTTTTTTTTTTTTTTTTTT

%%%%%%%%%%%%%%%%%%%%%%%%%%%%%%%%%%%%%%%%%%%%%%%%%%%%%%%%%%%%%%%%%%
\begin{figure}[ht]
\centering
%\hspace{-1.3cm}
\includegraphics[width= 7.5cm]{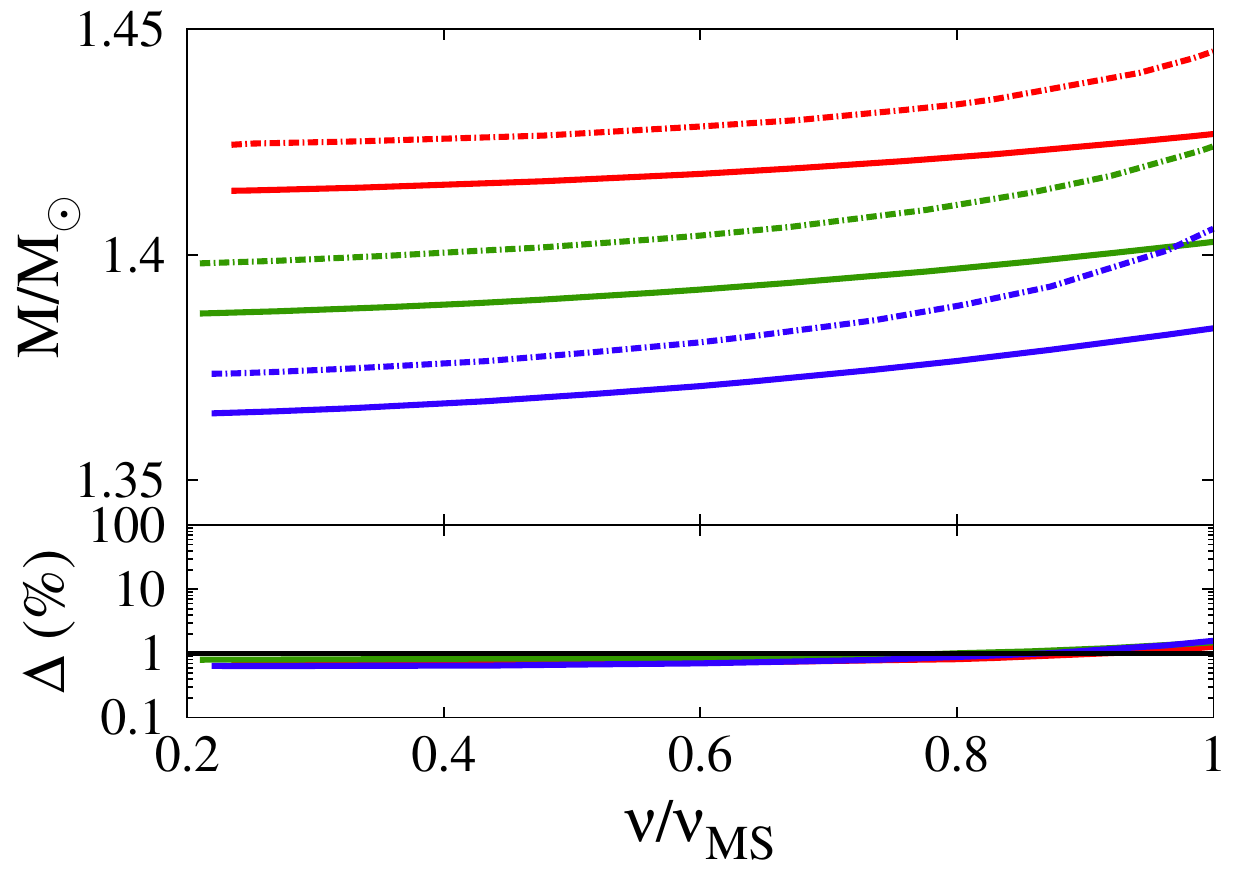}
\includegraphics[width= 7.5cm]{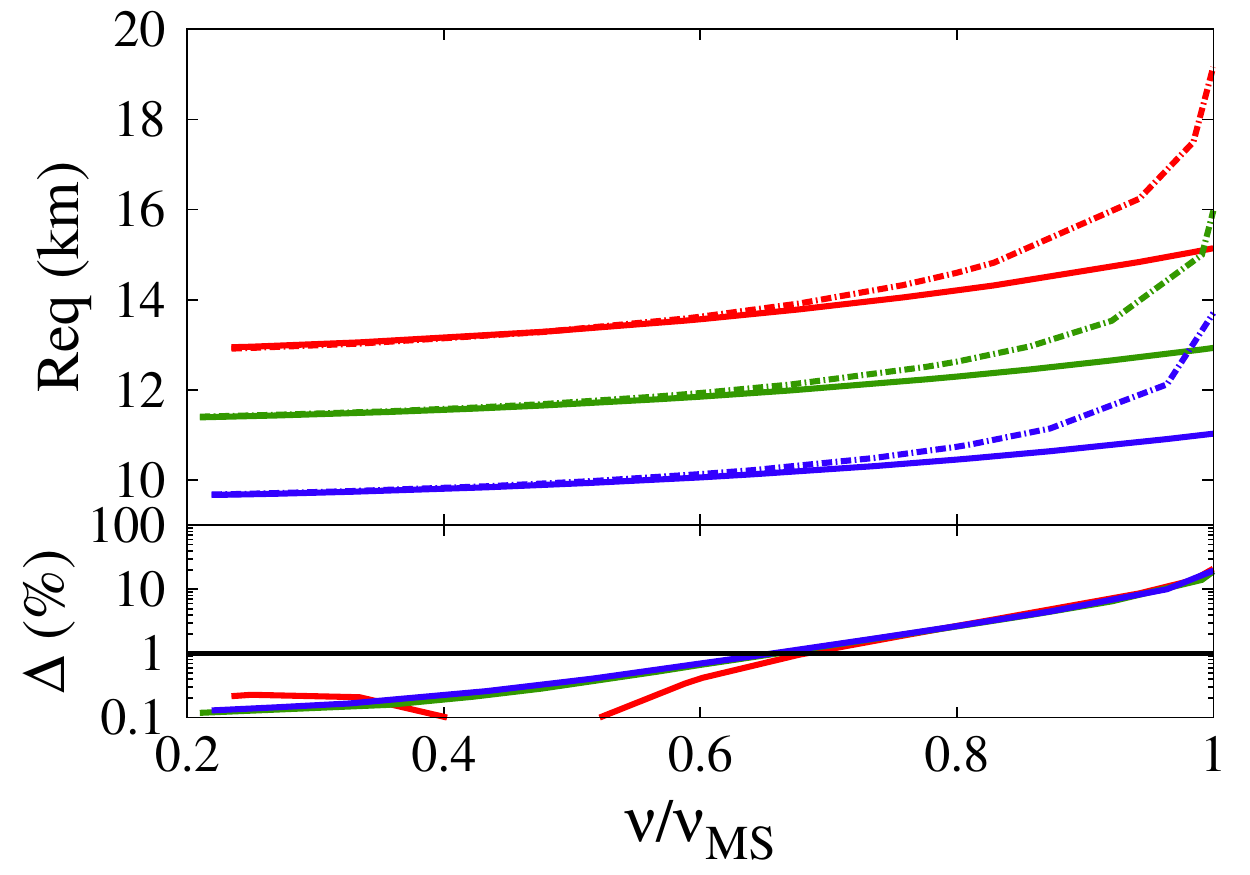}
\includegraphics[width= 7.5cm]{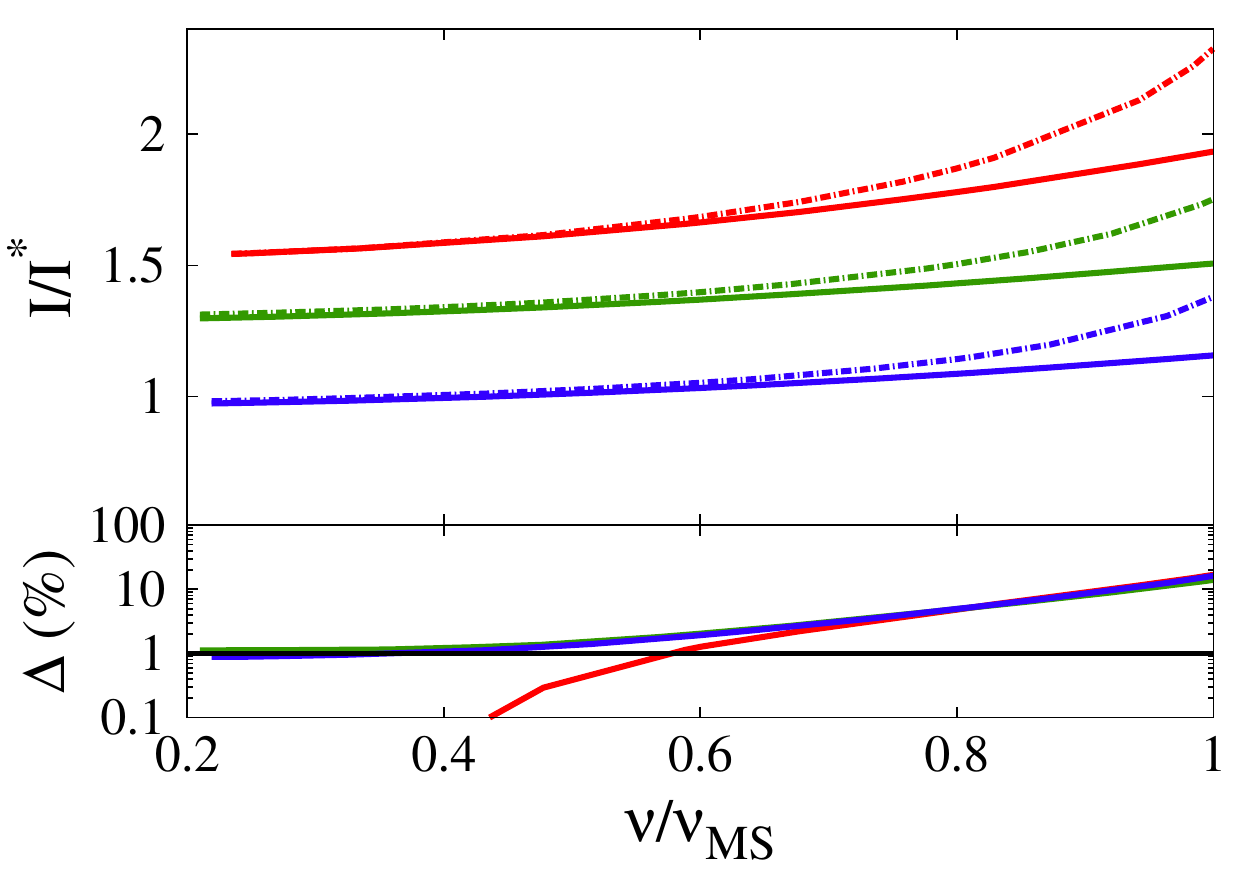}
\caption{(Color online) We plot the  mass, equatorial radius and moment of
inertia (normalized to $I^* = 10^{45}$ g$\times$cm$^2$),
as a function of the rotation rate
normalized to the mass-shedding  frequency.
Solid and dashed lines indicate
data computed by integrating Hartle's equations and using the RNS code,
respectively. All data refer to a baryonic mass $M_\tn{b}=1.55~M_\odot$
and the EoS \texttt{A}, \texttt{APR} and \texttt{g240}.
In the lower part of each panel we show
the relative difference between the data computed using the two
approaches.
}
\label{FIG9}
\end{figure}
%%%%%%%%%%%%%%%%%%%%%%%%%%%%%%%%%%%%%%%%%%%%%%%%%%%%%%%%%%%%%%%%%%

%TTTTTTTTTTTTTTTTTTTTTTTTTTTTTTTTTTTTTTTTTTTTTTTTTTTTTTTTTTTTTTT
\begin{table*}[h]
\centering
\begin{tabular}{cccc|ccc|ccc|ccc}
\hline
\hline
&\multicolumn{6}{c}{$M_\tn{b} = 1.55 M_{\odot}$}&\multicolumn{6}{|c}{$M_\tn{b}
=
2.2 M_{\odot}$}\\
\hline
& \texttt{g240} & \texttt{APR} & \texttt{A} & \texttt{g240} & \texttt{APR} & \texttt{A}
& \texttt{O} & \texttt{APR} & \texttt{AU} & \texttt{O} & \texttt{APR} & \texttt{AU}\\
\hline
$\nu/\nu_{\tn{ms}}$&& 0.6&&& 0.8&  &  & 0.6&&& 0.8&\\
\hline
$\nu$ (Hz) & 500& 653& 810 &  670& 871& 1080
& 650 & 793& 900& 860 & 1057& 1200\\
$\nu_{\tn{ms}}$ (Hz) & 833& 1089& 1350&  833& 1089& 1350
& 1069& 1321& 1495& 1069& 1321& 1495  \\
\hline
$\sqrt{M/R_{\tn{eq}}^3}$ (km$^{-1}$)  & 0.029 & 0.035& 0.045&  0.027& 0.032&
0.042
& 0.035 & 0.042& 0.048& 0.033& 0.039& 0.045 \\
$M/R_{\tn{eq}}$ & 0.15& 0.17& 0.20&  0.15 & 0.16 & 0.20
& 0.21 & 0.24& 0.26 & 0.21& 0.23& 0.25\\
\hline
$M^{\tn{Har}}/M_{\odot}$& 1.42& 1.39& 1.37&  1.42& 1.40 & 1.38
& 1.92& 1.89& 1.86 & 1.93 & 1.90 & 1.87\\
$M^{\tn{rns}}/M_{\odot}$  & 1.43& 1.40& 1.38&  1.43& 1.41& 1.39
& 1.94& 1.90& 1.87 & 1.95& 1.92& 1.89\\
$\Delta$ (\%)        & 1& 1& 1& 1& 1& 1& 1& 1& 1& 1& 1& 1\\
\hline
$R_{\tn{eq}}^{\tn{Har}}$ (km)  & 13.57& 11.85& 10.06&  14.20& 12.29 & 10.45
&13.32 & 11.53& 10.67& 13.81& 11.96 & 11.05\\
$R_{\tn{eq}}^{\tn{rns}}$ (km)  & 13.63& 11.93& 10.13&  14.60& 12.63& 10.74
& 13.42& 11.60& 10.74& 14.17& 12.25& 11.32\\
$\Delta$ (\%)        & < 1  & 1    & 1    &  3    & 3    & 3
& 1 & 1 & 1 & 3& 2& 2 \\
\hline
$I^{\tn{Har}}/I_*$        & 1.66 & 1.37 & 1.03 &  1.78 & 1.43& 1.09
& 2.70& 2.08& 1.85& 2.83& 2.18 & 1.93\\
$I^{\tn{rns}}/I_*$        & 1.69 & 1.40 & 1.05 &  1.87 & 1.51& 1.14
& 2.76& 2.12& 1.88& 2.99& 2.28 & 2.03 \\
$\Delta$ (\%)        & 2    & 2    & 2    &  5    & 5   & 4
& 2& 2& 2& 5& 4& 5 \\
\hline
\hline
\end{tabular}
\caption{ In this table we compare the values of the average density
$\propto\sqrt{M/R_{\tn{eq}}^3}$, compactness $M/R_{\tn{eq}}$, equatorial
radius $R_{\tn{eq}}$ and moment of inertia $I$ (normalized to $I^* = 10^{45}$
g$\times$cm$^2$),
found using Hartle's procedure and the RNS code. Data refer to two
values of the baryonic mass,
$M_\tn{b} = 1.55, 2.2 M_{\odot}$, and different EoS appropriate to describe
cold stars with that mass. For each quantity we also show the relative
difference  between RNS and Hartle's results. The comparison is done for
two values of the rotation rate, i.e. $\nu/\nu_{\tn{ms}}=0.6$ and
$\nu/\nu_{\tn{ms}}=0.8$.}
\label{table7}
\end{table*}
%TTTTTTTTTTTTTTTTTTTTTTTTTTTTTTTTTTTTTTTTTTTTTTTTTTTTTTTTTTTTTTTTTTTTTTTTTTT
%
The stellar parameters of the non rotating configurations
are given in Table~\ref{table6}.
The results of the Hartle-versus-RNS comparison are given in
Fig.~\ref{FIG9} and in Table~\ref{table7}.  In the three panels of
Fig.~\ref{FIG9} we compare mass, equatorial radius and moment of
inertia computed by integrating Hartle's equations
(solid lines) and using the RNS code (dashed lines).  All quantities are
plotted  versus the rotation rate normalized to the mass-shedding limit
rate (computed with RNS), and are given for a star with
baryonic mass $M_\tn{b}=1.55~M_\odot$
and  EoS, \texttt{A}, \texttt{APR} and \texttt{g240}.
In the lower part of each panel we also show
the percentual difference between Hartle's and RNS results.

In Table~\ref{table7} we tabulate  $M, R_{\tn{eq}}, I$ for two values of
the baryonic mass $M_\tn{b}=(1.55, 2.2)~M_\odot$ and two values of the
rotation rate, $\nu= (0.6,0.8)~\nu_{\tn{ms}}$. The superscripts `Har' and
`rns' indicate that the values have been obtained using the Hartle
procedure and the RNS code, respectively. For each quantity we also
provide the relative difference between the two approaches. In addition,
we show the average density, $\sqrt{M/R_{\tn{eq}}^3}$ and the compactness
$M/R_{\tn{eq}}$ of the star.

Fig.~\ref{FIG9} and Table~\ref{table7}
show that the relative difference between Hartle's and RNS's results is
\begin{itemize}
\item
less than $1\%$ for the gravitational mass for rotation rates $\lesssim
0.9 \nu_{\tn{ms}}$, where $\nu_{\tn{ms}}$ is the mass-shedding rotation rate
evaluated with RNS, for all considered EoS;
\item
less than $3\%$ for the equatorial radius up to $\lesssim
0.8 \nu_{\tn{ms}}$;
\item
less than $5\%$ for the moment of inertia up to  $\lesssim
0.8 \nu_{\tn{ms}}$;
\end{itemize}

We have checked that our results are independent of the resolution.
Doubling the number of grid points, all computed  values change by
less than 0.5\%.
%, except the quadrupole which can be affected by a $\sim 1
%\%$ change. This does not affect our conclusions.

A similar comparison between perturbative and fully non-linear codes has 
been done in \cite{Berti05}, by using Hartle's procedure at second order
in the rotation rate, and the RNS code. 
They were focussed on the evaluation of the innermost stable circular
orbit  and of the quadrupole
moment which, however, was affected by the RNS error mentioned in
Section~\ref{secIII}.
 
We also mention that in \cite{Yagi:2014bxa} Hartle's perturbative
approach has been extended up to fourth order in the rotation rate, increasing the
agreement with fully relativistic computations. However, 
a comparison with their results is not straightforward, because
they are focused on the determination of universal relations for  
higher order multipole moments, and this is 
beyond the scope of this paper.

\end{document}